\documentclass{ws-ijmpc}

\usepackage{subfigure}
\usepackage[super,numbers,sort&compress]{natbib} %Ö§³Ö¶àÎÄÏ×¼¯ÖÐÒýÓÃ
\makeatletter %µÃµ½Í¼ÐαêÌ⣬Ëƺõ±¾ÎÄûÓõ½
\newcommand\figcaption{\def\@captype{figure}\caption}
\newcommand\tabcaption{\def\@captype{table}\caption}
\makeatother

\begin{document}

\markboth{J. Fang et al.}
{A Two-Dimensional CA Traffic Model with Dynamic Route Choices Between Residence and Workplace}

\catchline{}{}{}{}{}
%%%%%%%%%%%%%%%%%%%%%%%%%%%%%%%%%%%%%%%%%%%%%%%%%%%%%%%%%%%%%%%%%%%%

\title{A TWO-DIMENSIONAL CA TRAFFIC MODEL WITH DYNAMIC ROUTE CHOICES BETWEEN RESIDENCE AND WORKPLACE}

\author{JUN FANG\footnote{Corresponding author.}}

\address{Department of Computer Science and Technology\\
Tsinghua University, Beijing 100084, P. R. China\\
fangjun06@mails.tsinghua.edu.cn}

\author{JING SHI}

\address{Institute of Transportation Engineering\\
Tsinghua University, Beijing 100084, P. R. China\\
jingshi@tsinghua.edu.cn}

\author{XI-QUN CHEN}

\address{Department of Civil Engineering\\
Tsinghua University, Beijing, 100084, P. R. China\\
chenxq04@mails.tsinghua.edu.cn}

\author{ZHENG QIN}

\address{Department of Computer Science and Technology \\
Tsinghua University, Beijing 100084, P. R. China\\
qingzh@tsinghua.edu.cn}

\maketitle

%\begin{history}
%\received{Day Month Year}
%\revised{Day Month Year}
%\end{history}

\begin{abstract}
The Biham, Middleton and Levine (BML) model is extended to describe dynamic route choices between the residence and workplace in cities. The traffic dynamic in the city with a single workplace is studied from the velocity diagram, arrival time probability distribution, destination arrival rate and convergence time. The city with double workplaces is also investigated to compared with a single workplace within the framework of four modes of urban growth. The transitional region is found in the velocity diagrams where the system undergoes a continuous transition from a moving phase to a completely jamming phase. We perform a finite-size scaling analysis of the critical density from a statistical point of view and the order parameter of this jamming transition is estimated. It is also found that statistical properties of urban traffic are greatly influenced by the urban area, workplace area and urban layout.

\end{abstract}

\ccode{PACS Nos.: 89.40.-a, 05.70.Fh, 05.40.-a.}

\section{Introduction}

In 1992, O. Biham, A. A. Middleton and D. Levine proposed a two-dimensional CA traffic model (BML) to investigate the phase transition and self-organization effect in the traffic flow of two dimensions.\cite{Biham1992} The model is defined at a square lattice with $N\times N$ sites. Within a given time step each site can be held at most by one car with three states: moving upward, moving rightward and standing still. The car movements are controlled by synchronous traffic lights that allow right moving only in even time steps and up moving in odd time steps. At their turn, every car moves to its neighbor site ahead whenever the site remains vacant. A sharp jamming transition was found between the free  phase at low density and the completely jamming phase at high density. Recently, many generalizations and extensions of the BML model were proposed to investigate various realistic traffic factors in cities, such as the anisotropy effect, independent turning and jam-avoiding turning, signal control, roads between successive crossings, etc.\cite{Chowdhury2000, Helbing2001, Nagatani1993, Martinez1995, Nagatani1995, Torok1996, Benyoussef2003, Belbasi2008, Spyropoulou2007, Toledo2004, Chowdhury1999, Brockfeld2001, Shi2007}

Moussa incorporated the origin-destination effect of drivers trips in the BML model to study the traffic demand in cities.\cite{Moussa2007} To characterize the trip, Moussa used three different O-D distance probability distribution (ODDPD): exponential, uniform, and power-law. Moussa investigated the effect of the ODDPD on the traffic flow and the arrival times of travelers. Moussa subsequently studied the evacuation processes of cars in cities using above model.\cite{Moussa2009} It was found that the ODDPD has a great effect on the evacuation processes and the evacuation time diverges in obedience to the power law. In-nami et al. introduced a similar model to investigate the phase transitions on a decorated square-lattice where the square-lattice point and the decorated site denoted the intersection and road, respectively.\cite{In-nami2007} In their model, a car has a finite deterministic path between the origin and the destination, which is assigned to the car from the beginning.

%There are some similarities between Ref.~\refcite{Moussa2007} and Ref.~\refcite{In-nami2007}. But in Ref.~\refcite{In-nami2007} when a car reach its destination, it is deleted and a new car is created at a vacant site that is chosen randomly. However, in Ref.~\refcite{Moussa2007} the car exists after arriving the destination and a new destination is selected for it according to three different O-D distance probability distribution (ODDPD): exponential, uniform, and power-law.

%We develop the BML model to incorporate the residence-workplace effect of driver trips on the traffic in cities. We choose a block of square lattice to be the workplace in a city and take the surrounding as the residence. Each car is associated with a site of the residence without overlaps as the origin point and its destination point is randomly chosen from the workplace. For each trip, a shortest path is found through the lattice. After arriving the destination assigned for the car, it will disappear (taken to the underground garage). Dynamic phase transitions for different sizes of the city and the workplace are studied numerically.

Huang simulated the city traffic in rush hours in a single-lane roadway with open boundaries using the Nagel-Schreckenberg traffic model.\cite{Huang2007} The influence of the traffic light control to the traffic flow and travel time was studied. In the morning rush hours, downtown parking lots are taken as the destinations for cars moving from suburban. When a car arrives at its parking lot, it sinks immediately. In the afternoon rush hours, the traffic is in the opposite direction from the morning rush hours and original sinks become the sources for cars moving out of the downtown. In the present paper, we extend the scenario in road from one dimension to two dimensions and introduce the dynamic route choices into car movements within the framework of BML traffic model. We find that statistical properties of urban traffic are greatly influenced by the urban area, workplace area and urban layout.

This paper is organized as follows: In Sec.~\ref{model}, we elaborate our model for city traffic with the dynamic route choices between the residence and workplace. In Sec.~\ref{simulation}, dynamic phase transitions for different sizes of the city and workplace are studied numerically. We also investigate the traffic dynamic in the city with double workplaces for comparison with a single workplace from the velocity diagram, arrival time probability distribution, destination arrival rate and convergence time. Finally, we draw some conclusions in Sec.~\ref{conclusion}.

\section{Model} \label{model}
In the original BML traffic model, cars will keep on moving along the vertical and horizontal lanes without any target sites if traffic lights permit them to pass. In this study, we extend the BML model to describe dynamic route choices between the residence and workplace on the traffic in cities. We plan to simulate the scenario that office workers drive toward workplaces from their residences every morning. The present model includes four modules:
\begin{figure}[htb]
\begin{minipage}[t]{0.45\linewidth}
\centering
\includegraphics[width=2.2in]{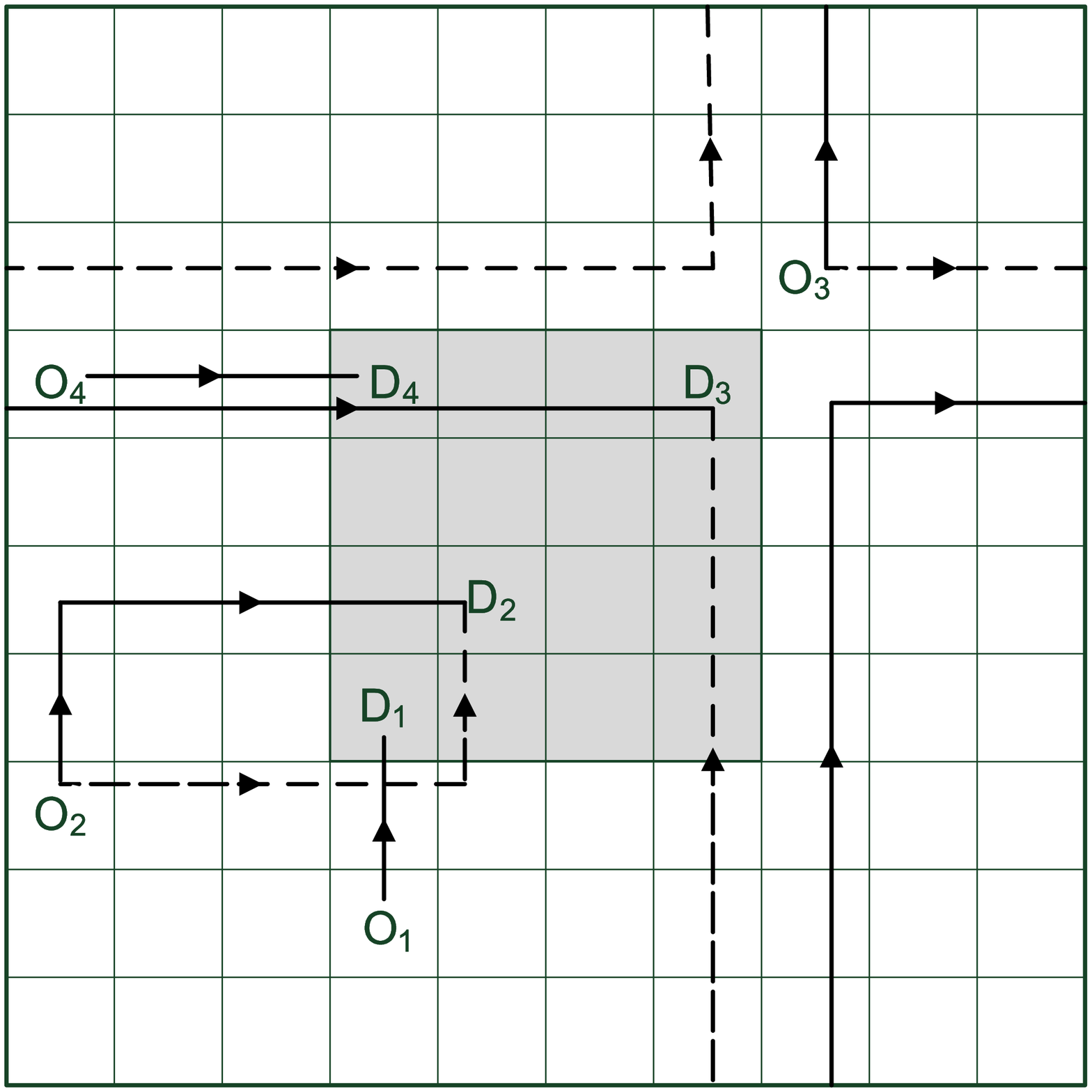}
\caption{Illustration of the dynamic route choices in a square lattice with the periodic boundary. There are four cars moving from their origins (see O$_{1}$ O$_{2}$, O$_{3}$, O$_{4}$) to destinations (see D$_{1}$ D$_{2}$, D$_{3}$, D$_{4}$), which is similar to Ref.~11. In our model cars can only move rightward or upward. If its destination is at the left of the origin, or on the low side (see O$_{3}$ and D$_{3}$), the car must pull off the lattice and then enter the lattice again from the opposite side (see along the dash or solid line linking O$_{3}$ and D$_{3}$).}
\label{route}
\end{minipage}
\hfill
\begin{minipage}[t]{0.45\linewidth}
\centering
\includegraphics[width=2.2in]{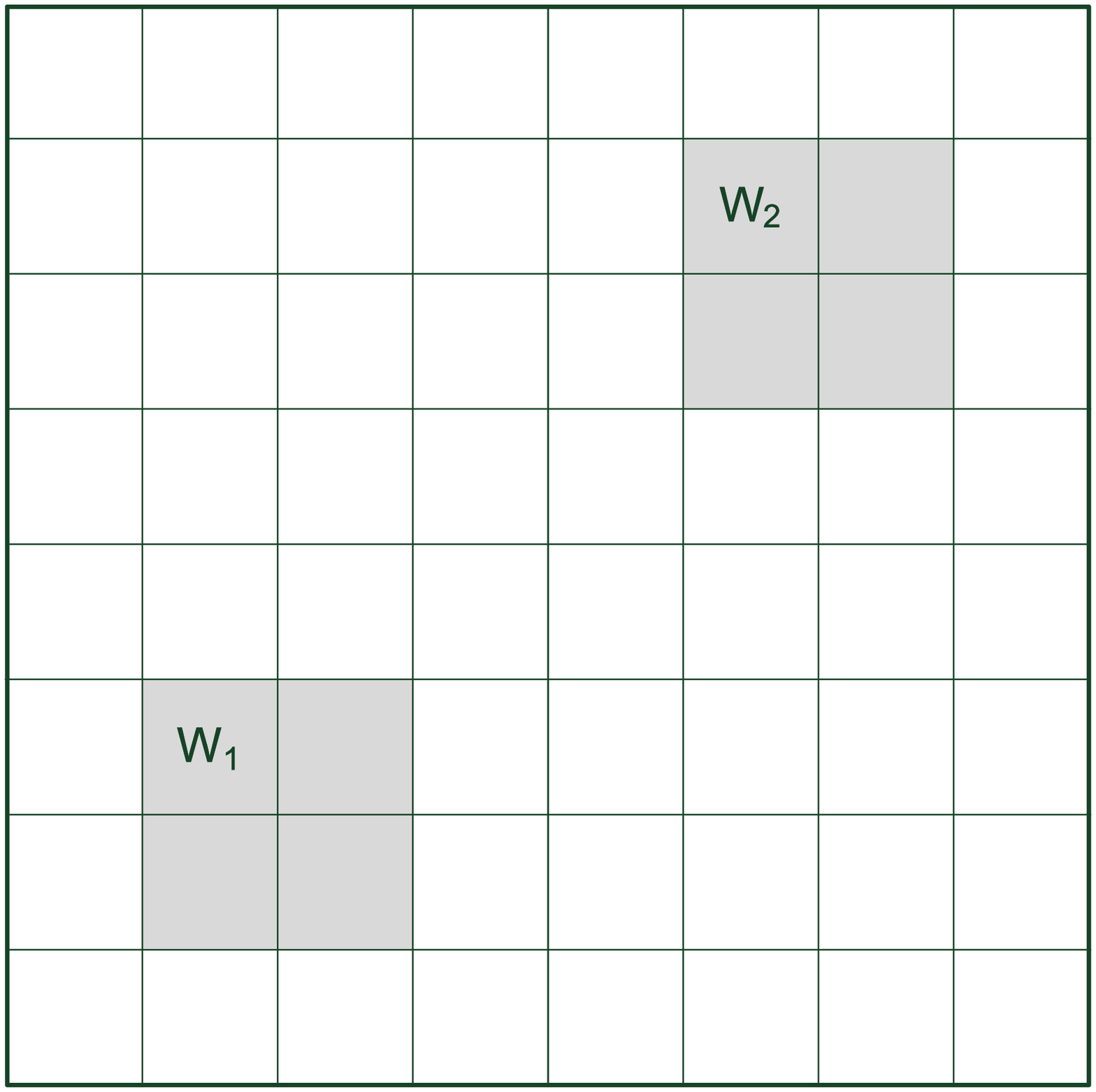}
\caption{Illustration of the square lattice (city) of size $L=8$ with double workplaces of size $M=2$. The two workplaces, denoted by $\textrm{W}_{1}$ and $\textrm{W}_{2}$ in gray areas, are arranged uniformly in the city with equal vertical and horizontal distance (with the periodic boundary).}
\label{two}
\end{minipage}
\end{figure}

\begin{romanlist}[(i)]
\item Regional division: In our model, the city is divided into two regions: the residence and the workplace. The workplace is a square of the plane lying in the middle of the city (See the gray area in Fig.~\ref{route}),\footnote{For the periodic boundary used in our model, the position of the workplace has no effect on simulation results. Just for the convenience of demonstration, the workplace is put in the center of the city.} around which is the residence (See the white area in Fig.~\ref{route}).
\item Initialization of simulation: Each car is assigned a site  randomly without overlaps from the residence as its origin, and a site randomly from the workplace as its destination. The same destination for different cars in the workplace is allowed. The origin-destination information is assigned to every car at the beginning of simulations and all cars start out simultaneously on each way to their destinations.
\item Dynamic route choices: In our model, cars select their shortest pathes with a simple method similar to Ref.~\refcite{Moussa2007} (or see Fig.1). There are two pathes a car can select to go to its destination. The car can move upward firstly and turn to right when it reaches the site belonging to the lane of the destination site. Then it keeps on going until it reaches the destination (see the solid arrow line between O$_{2}$ and D$_{2}$ in Fig.~\ref{route}). On the other side, the car may choose another path (see the dash arrow line between O$_{2}$ and D$_{2}$ in Fig.~\ref{route}). For the periodic boundary, the route may go out of bound and then go in from the opposite side (e.g. from O$_{3}$ to D$_{3}$). Every car can select alternatively from the two types of pathes. In particular, if the origin and destination site are in the same column (or lane), the car has no choice but to move upward (or rightward) (e.g. from O$_{1}$ to D$_{1}$ or from O$_{4}$ to D$_{4}$). Counting these cars out, whose origin and destination sites are in the same column or lane, half of cars are assigned randomly to move upward and the other half to move rightward on departure to avoid the anisotropy effect.

\item Placement of cars after arriving their destinations: We assume that in the workplace, there are enough underground garages to accommodate all cars coming from the residence. Every site of workplace can hold an infinite number of cars. When a car arrives at a destination site, it disappears immediately without spending time on parking.
\end{romanlist}

\section{Simulations and Results}
\label{simulation}
We carry out simulations in a square lattice (city) of size $L$ the workplace of size $M$. We put a number $N$ of cars randomly all over the residence and leave the workplace vacant. The density $\rho$ referred to in this paper is defined as the initial density of cars in the residence, which is denoted as $\rho=N/(L^{2}-M^{2})$.

We investigate four typical city sizes $L=\{64,128,256,512\}$ and two types of city layouts, i.e. the city with a single workplace (C-1) and the city with double workplaces (C-2). We only consider that the two workplaces are distributed evenly in the city, which is shown in Fig.~\ref{two}. To avoid the anisotropy effect, the number of cars moving horizontally at the first time step equals that of cars moving vertically.

The simulation terminates under three criteria:
\begin{romanlist}[(i)]
\item If all cars disappear, i.e. every car has arrived at the destination and no one is still on the way, the simulation ends spontaneously.\label{cri:i}
\item If all cars remaining in the city are jammed with traffic and no one could continue to move, we terminate the simulation.\label{cri:ii}
\item If the simulation lasts too long and can't still meet the criterion i or ii, it will be terminated forcibly. We make statistics utilizing data generated from the last 100 time steps. Limited by our computing power, the upper simulation time is set to 100,000 steps. In the following text, one time step is referred as one signal phase of moving rightward plus one signal phase of moving upward.\label{cri:iii}
\end{romanlist}

We mainly investigate four typical scenarios about city growing:

\begin{romanlist}[(i)]
\item Scenario-1: the area ratio ($r$) of the workplace to the city remains at $10\%$.
\item Scenario-2: the area ratio ($r$) of the workplace to the city remains at $20\%$.
\item Scenario-3: the area of workplace remains constant during the city expansion. We set size $M=20$ for the workplace of C-1 and $M=14.14$ for every workplace of C-2.
\item Scenario-4: the workplace narrows down to include only one site, which is the extreme case of Scenario-3.
\end{romanlist}

\subsection{Velocity diagrams}
\label{vd}
The velocity of each car can be either 1 or 0. The average velocity $\langle v_{t}\rangle$ of cars at time $t$ is defined as the ratio of moving cars to all cars leaving in the city at time $t$. We mainly investigate the ensemble average velocity (denoted by $\langle v\rangle$) by repeated many simulations under the same conditions and averaged over the asymptotic configurations. If the simulation terminates by the criterion (\ref{cri:i}), i.e. all cars arrive at their destinations within the given time, $\langle v\rangle$ is set to one. If only a part of cars arrive at their workplaces and the rest get stuck in traffic jam (criterion (\ref{cri:ii})), $\langle v\rangle$ is zeroed. If the given time (100,000 time steps) expires and some cars are still running, $\langle v\rangle$ is averaged out by taking $\langle v_{t}\rangle$ of the last 100 time steps.

\begin{figure}[htb]
\subfigure[]{
\label{v64:a} %% label for first subfigure
\begin{minipage}[b]{0.5\textwidth}
\centering
\includegraphics[width=2.7in]{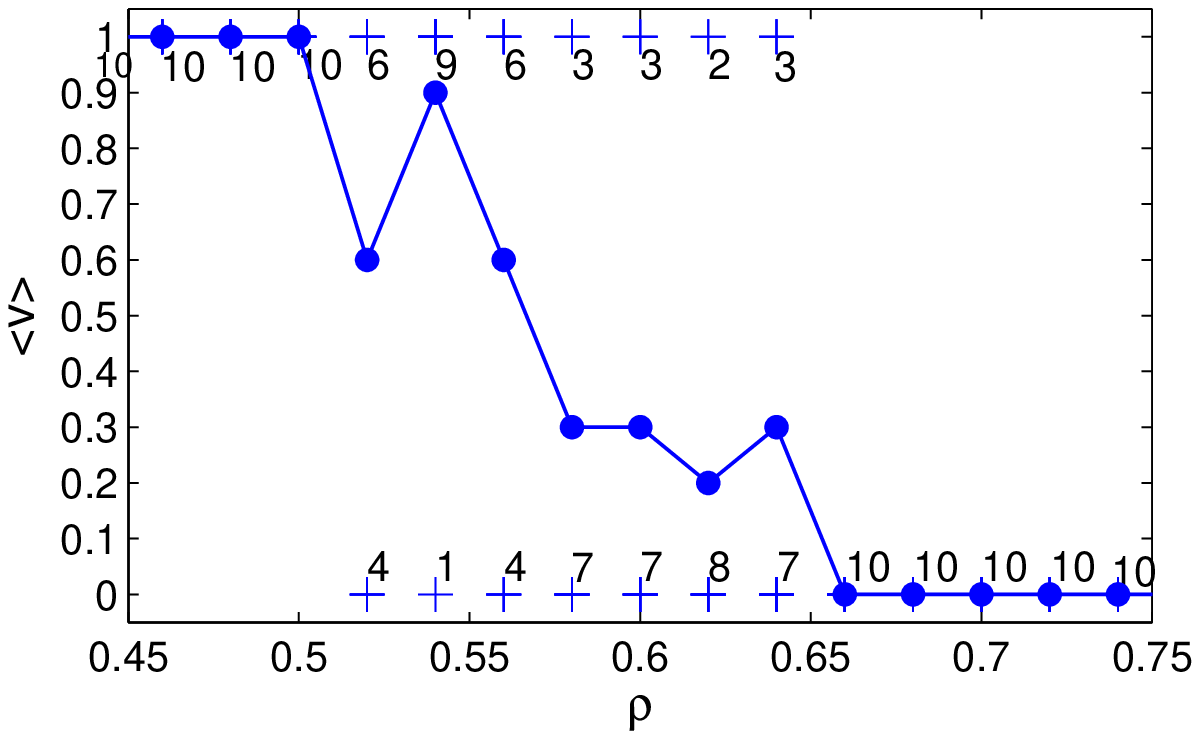}
\end{minipage}}%
\subfigure[]{
\label{v64:b} %% label for second subfigure
\begin{minipage}[b]{0.5\textwidth}
\centering
\includegraphics[width=2.7in]{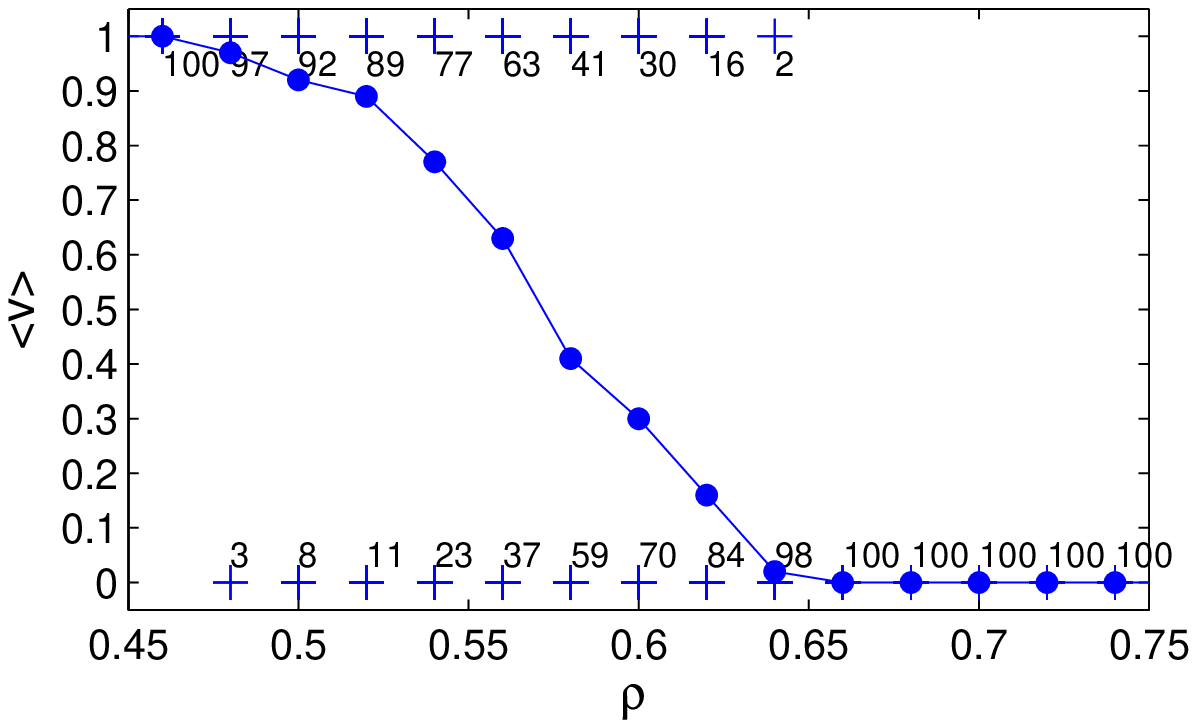}
\end{minipage}}
\caption{The polyline shape of average velocity at the end of simulation is influenced by the number of samples in every density. In every density, (a) $\langle v\rangle$  is averaged out from 10 samples; (b) $\langle v\rangle$  is averaged out from 100 samples. We subscribe (or superscribe) each plus sign with a number indicating the times of $\langle v\rangle=1$ (or $\langle v\rangle=0$). Every circle is calculated by taking the mean of the times of $\langle v\rangle=1$ and $\langle v\rangle=0$ in every density. The surveyed density increases with the step length of 0.02. We find in numerous simulations that the fluctuation of curve still cannot be eliminate  entirely even after adopting 100 samples (not shown in Fig.~\ref{v64}).
}
\label{v64} %% label for entire figure
\end{figure}

\begin{figure}[htb]
\subfigure[]{
\label{vr:a}
\begin{minipage}[]{0.5\textwidth}
\centering
\includegraphics[width=2.6in]{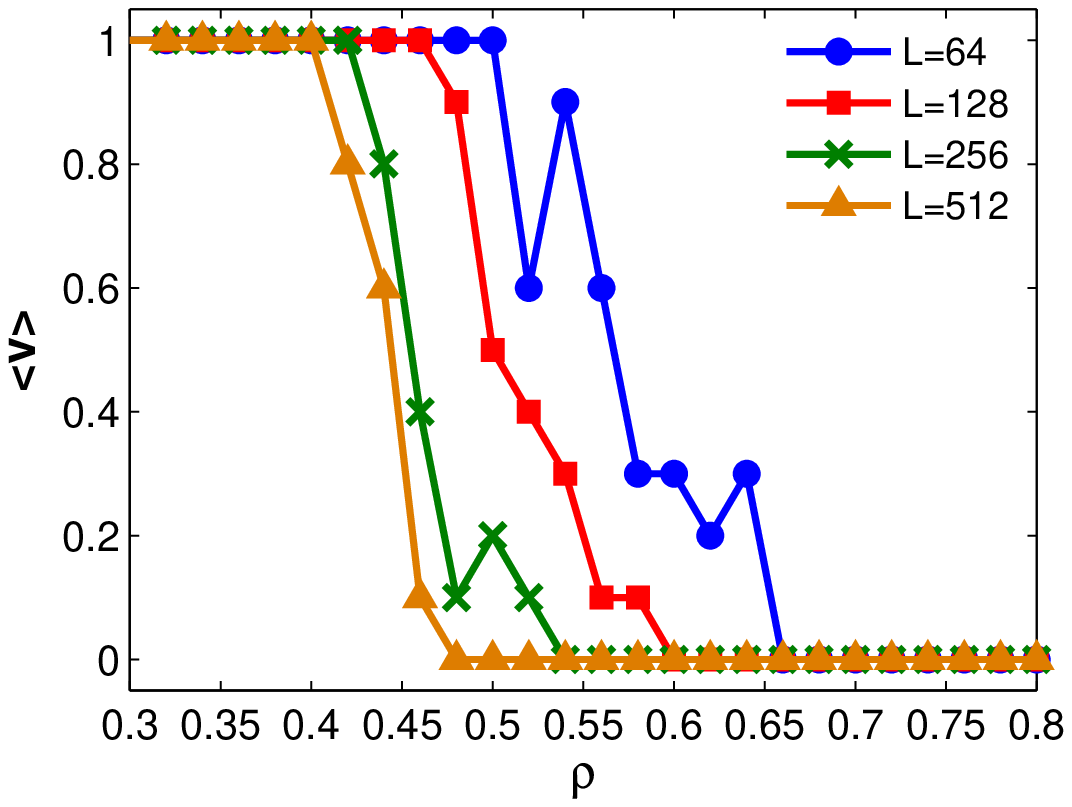}
\end{minipage}}%
\subfigure[]{
\label{vr:b}
\begin{minipage}[]{0.5\textwidth}
\centering
\includegraphics[width=2.6in]{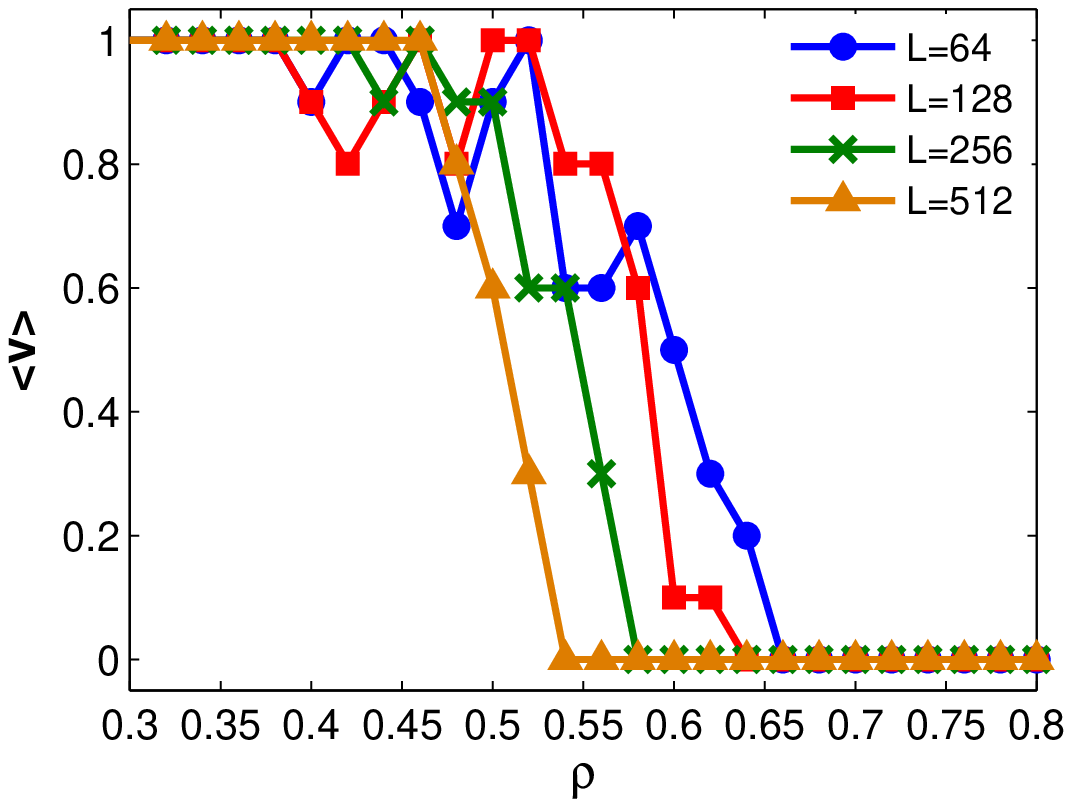}
\end{minipage}} \\[-10pt]
\subfigure[]{
\label{vr:c}
\begin{minipage}[]{0.5\textwidth}
\centering
\includegraphics[width=2.6in]{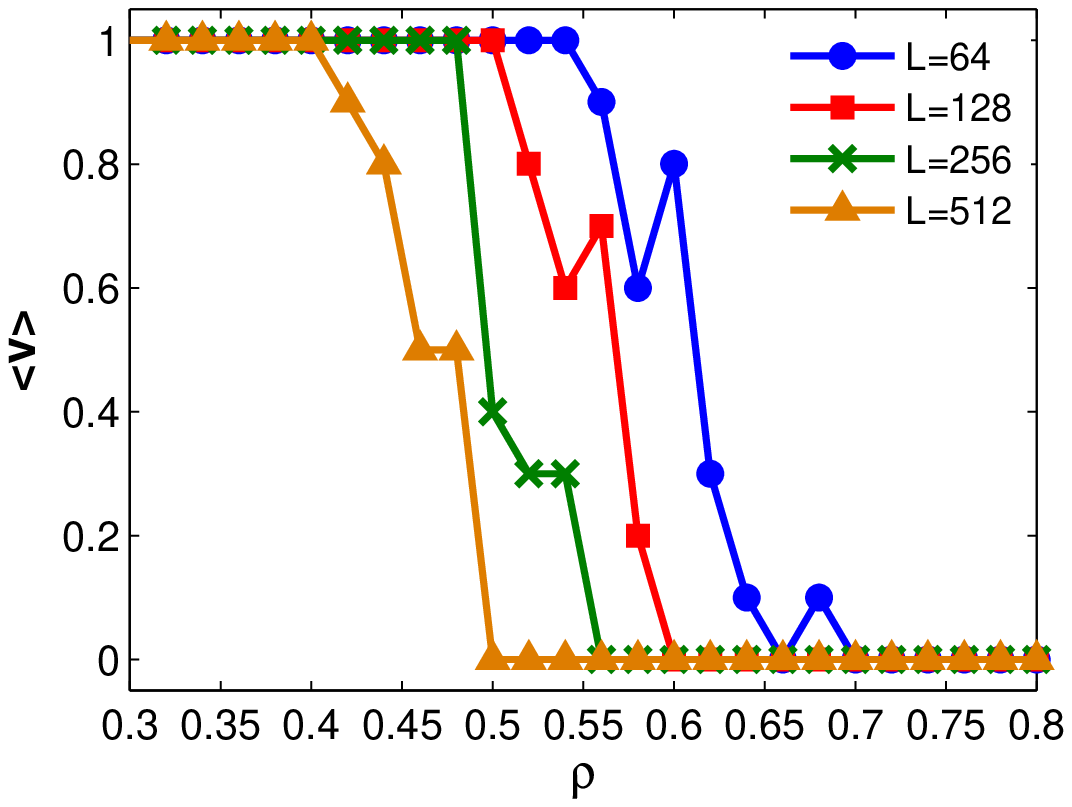}
\end{minipage}}%
\subfigure[]{
\label{vr:d}
\begin{minipage}[]{0.5\textwidth}
\centering
\includegraphics[width=2.6in]{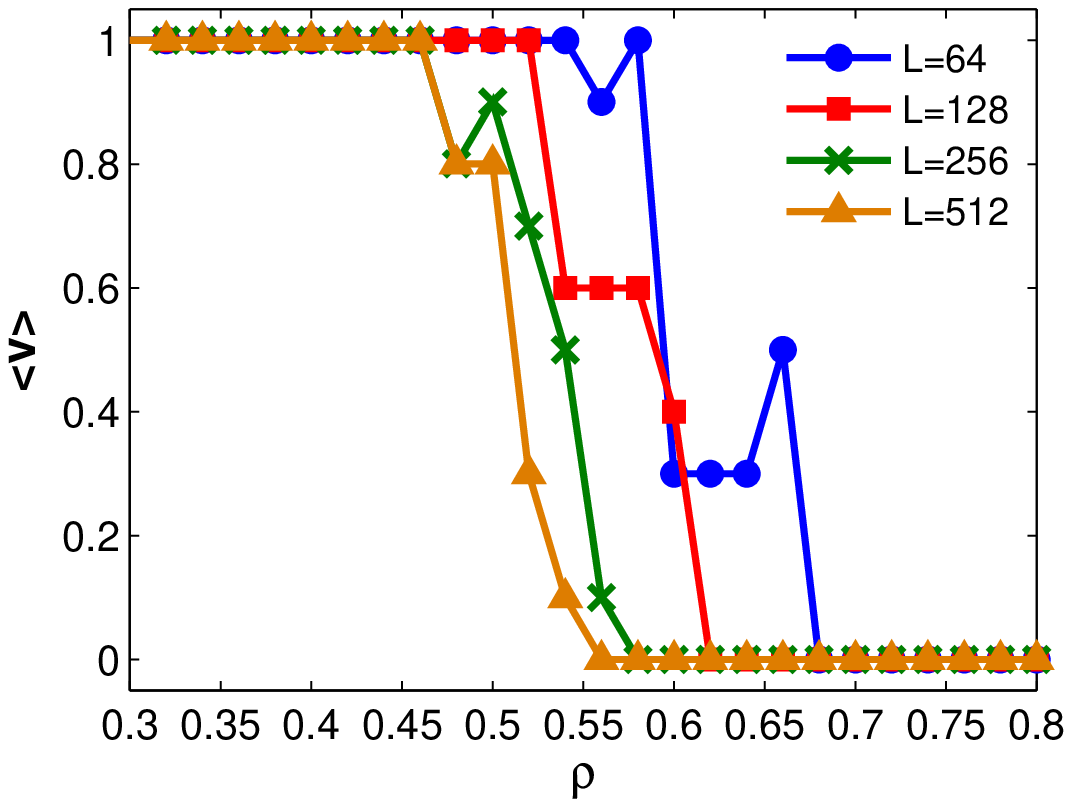}
\end{minipage}} \\[-10pt]
\subfigure[]{
\label{vr:e}
\begin{minipage}[]{0.5\textwidth}
\centering
\includegraphics[width=2.6in]{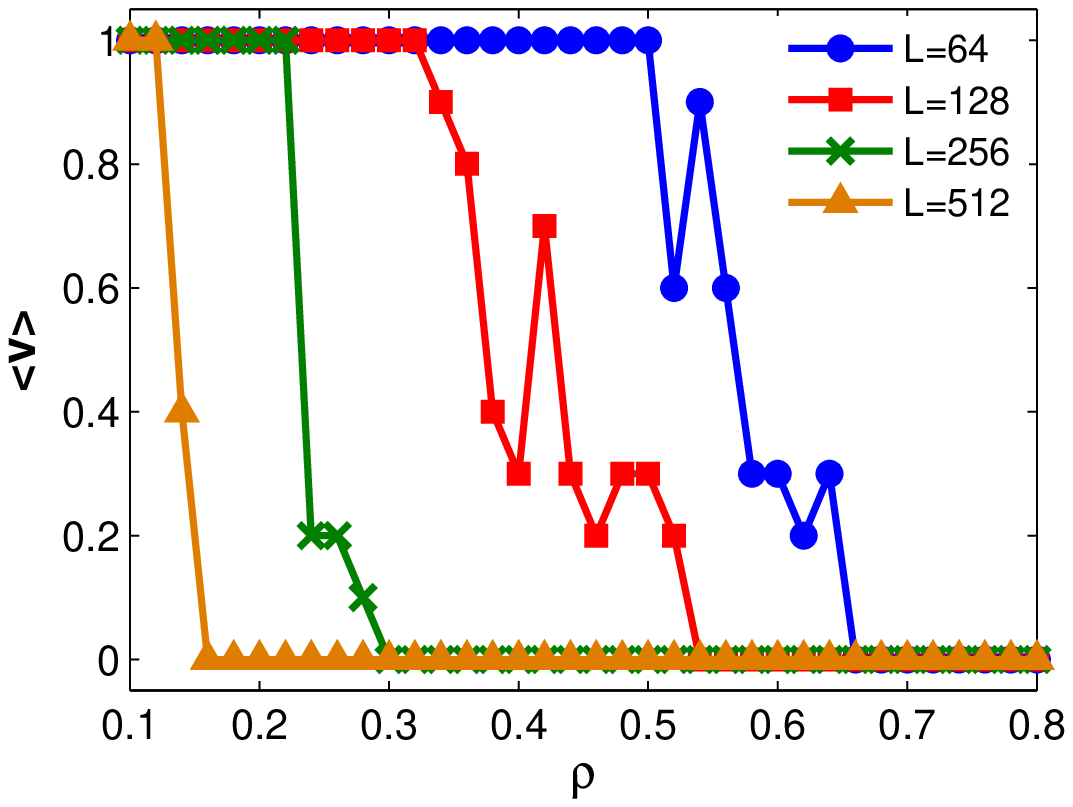}
\end{minipage}}%
\subfigure[]{
\label{vr:f}
\begin{minipage}[]{0.5\textwidth}
\centering
\includegraphics[width=2.6in]{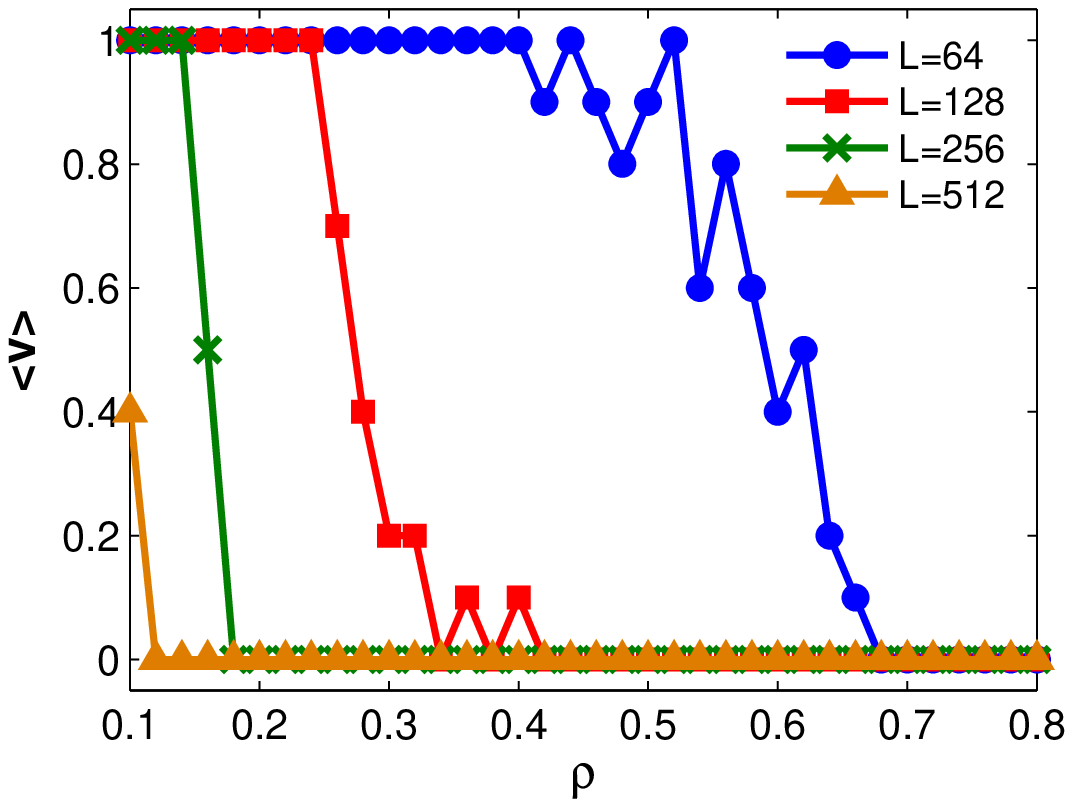}
\end{minipage}}
\caption{The average velocity ($\langle v\rangle$) versus density ($\rho$) in the city of (a) C-1 in the Scenario-1; (b) C-2 in the Scenario-1; (c) C-1 in the Scenario-2; (d) C-2 in the Scenario-2; (e) C-1 in the Scenario-3; (f) C-2 in the Scenario-3, where C-1 and C-2 represent the city with a single workplace and double workplaces, respectively. The surveyed density increases with the step length $0.02$. Every point of $\langle v\rangle$ is drawn by the mean value of ten implements on the corresponding density. }
\label{vr}
\end{figure}

In order to weaken the random effects of the simulation, ten implements are implemented in every density to calculate the average value of $\langle v\rangle$. Ten samples may not be enough to obtain $\langle v\rangle$ accurately. Comparing the velocity diagrams of different samples in every density in Fig.~\ref{v64}, we find that with the increase of samples, the trend of decreasing monotonically is obvious. We judge that the fluctuation and the nonmonotonic trend of the polyline in Fig.~\ref{v64}(a) is due to the randomness of car initial positions. However, adopting greater than 10 implements in every density is beyond our current computing power, especially for the large city size (e.g. $L=512$). We have to set 10 samples in every density as a compromise.

Fig.~\ref{vr} shows the average velocity ($\langle v\rangle$) as a function of the density ($\rho$) for the four different city sizes and two types of urban layouts. The velocity diagrams display that $\langle v\rangle$ decreases from one to zero gradually with the increase of density and vanished beyond the critical density $\rho_{c}$. There is a transitional region in the velocity diagrams where the system undergoes a continuous transition from a moving phase ($\langle v\rangle\neq0$) to a completely jamming phase ($\langle v\rangle=0$). Comparing different city sizes, we find $\rho_{c}$ keeps decreasing as $L$ increases coinciding with Ref.~\refcite{Biham1992}.
%However, we don't observe that as the lattice size increases the transition becomes sharper, which was reported in Ref.~\refcite{Biham1992}.

Comparing the velocity diagrams of C-1 and C-2 in the Scenario-1 in Fig.~\ref{vr}(a)-(b), we find that the critical density of C-2 is about 0.05 higher than that of C-1 except for $L=64$. The result indicates that the jam probability of the urban layout of double workplaces is less than that of a single workplace especially for large city. The curves of C-2 stand closer by one another than that of C-1, which reveals the C-2 is less insensitive to the finite size effect than C-1.

The velocity diagram in the Scenario-2 is shown in Fig.~\ref{vr}(c)-(d), which is very similar to the Scenario-1. The $\rho_{c}$ in the Scenario-2 is greater than Scenario-1 and the growth amplitude is larger for C-1 than C-2. This indicates the road capacity gain from the increment of the workplace area is more considerable for C-1 than C-2.

In Fig.~\ref{vr}(e)-(f), it is obvious that the traffic is terrible for both layouts in huge cities. The $\rho_{c}$ in the Scenario-3 is extremely lower than Scenario-1,2 when $L=128,256,512$ whether for C-1 or C-2. Hence it would be unwise to enlarge the city while remain the area of workplace or city center constant. When the city extends to $L=128,256,512$, the $\rho_{c}$ of C-2 is lower than C-1, which is not discovered in the Scenario-1,2. Why this could happen is still unknown.

\subsubsection{The 4th scenario: the workplace narrows down to include one site}
\label{sce-4}
\begin{figure}[htb]
\subfigure[]{
\label{ov:a} %% label for first subfigure
\begin{minipage}[b]{0.5\textwidth}
\centering
\includegraphics[width=2in]{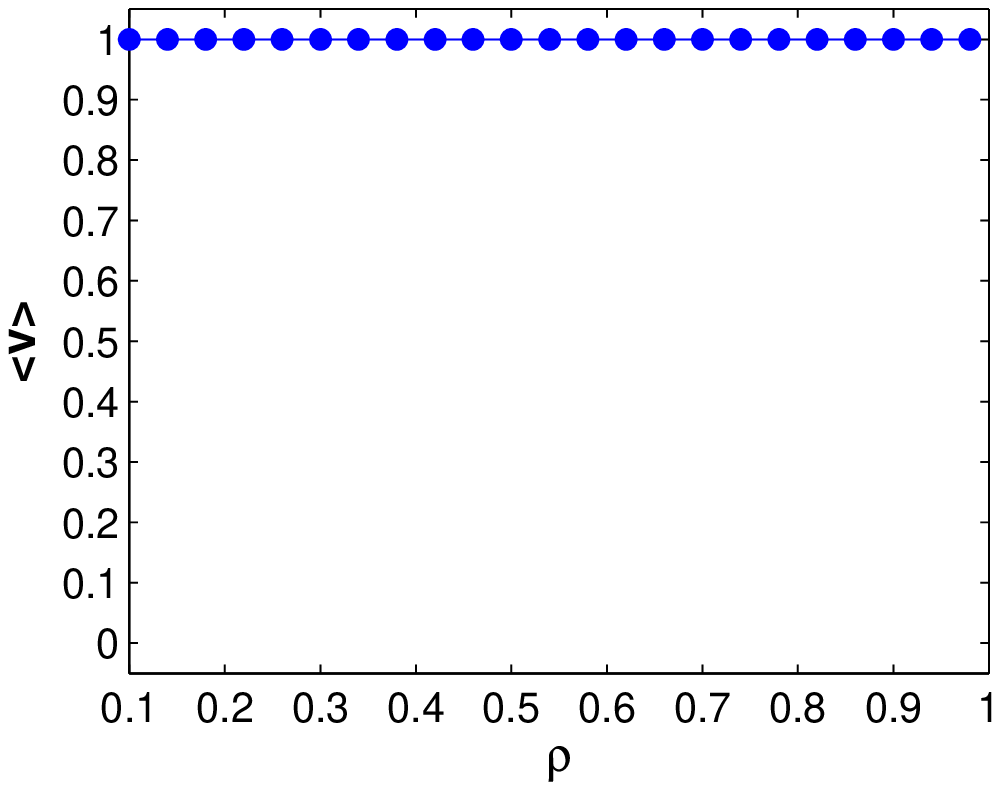}
\end{minipage}}%
\subfigure[]{
\label{ov:b} %% label for second subfigure
\begin{minipage}[b]{0.5\textwidth}
\centering
\includegraphics[width=2in]{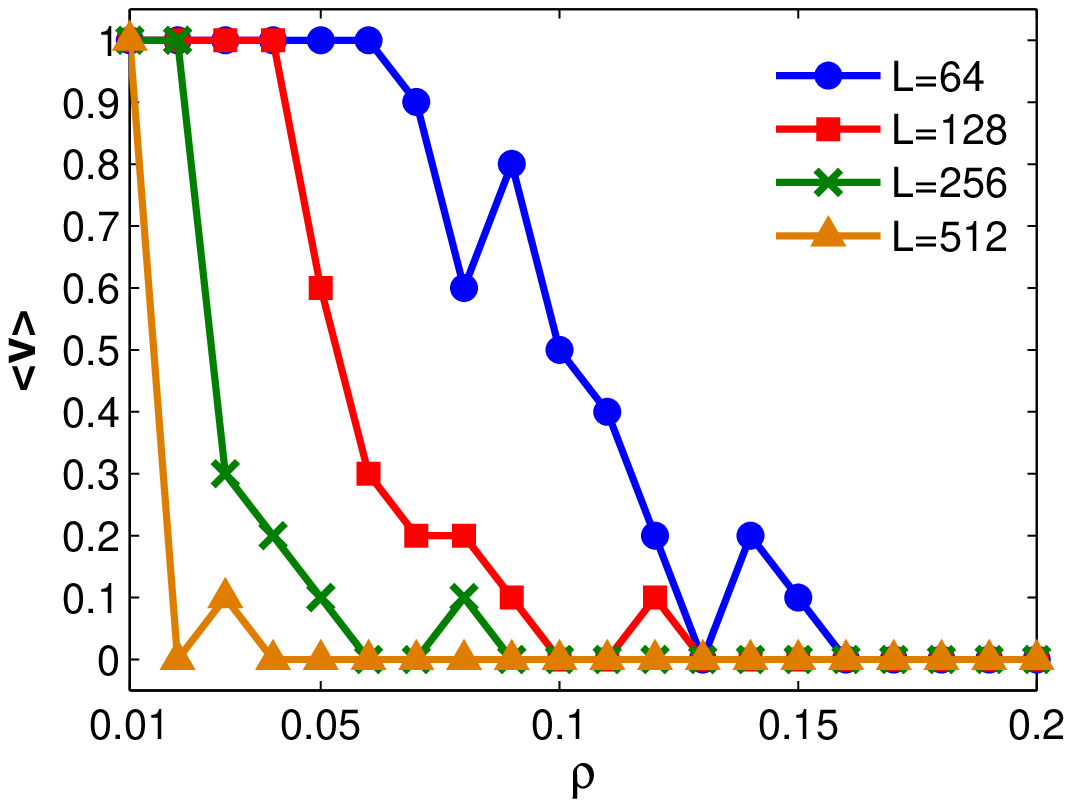}
\end{minipage}}
\caption{The velocity diagrams for the (a) a single workplace with only one site; (b) double workplaces each of which has only one site. In Fig.~\ref{ov:a}, $\langle v\rangle$ remains one for city sizes $L={64, 128, 256, 512}$ in any density.}
\label{ov} %% label for entire figure
\end{figure}

\begin{figure}[htb]
\subfigure[]{
\label{o:a}
\begin{minipage}[]{0.33\textwidth}
\centering
\includegraphics[width=1.7in]{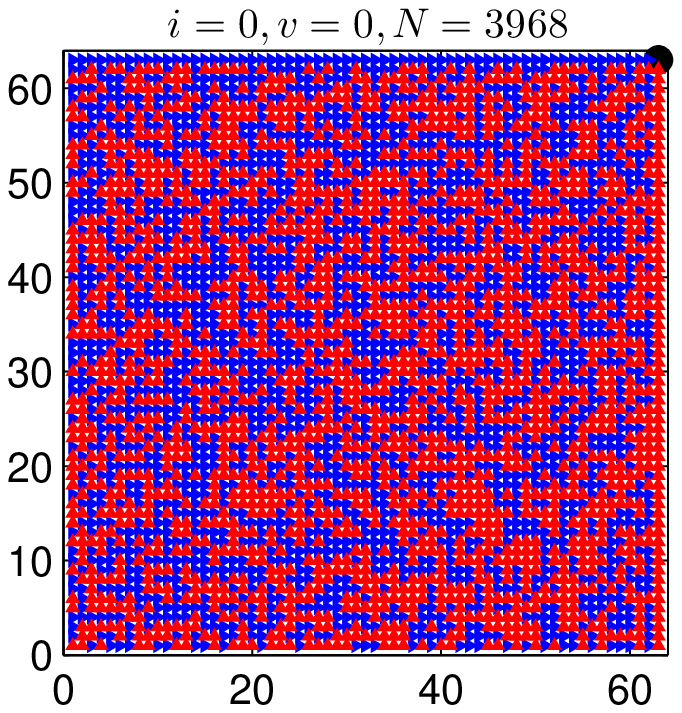}
\end{minipage}}%
\subfigure[]{
\label{o:b}
\begin{minipage}[]{0.33\textwidth}
\centering
\includegraphics[width=1.7in]{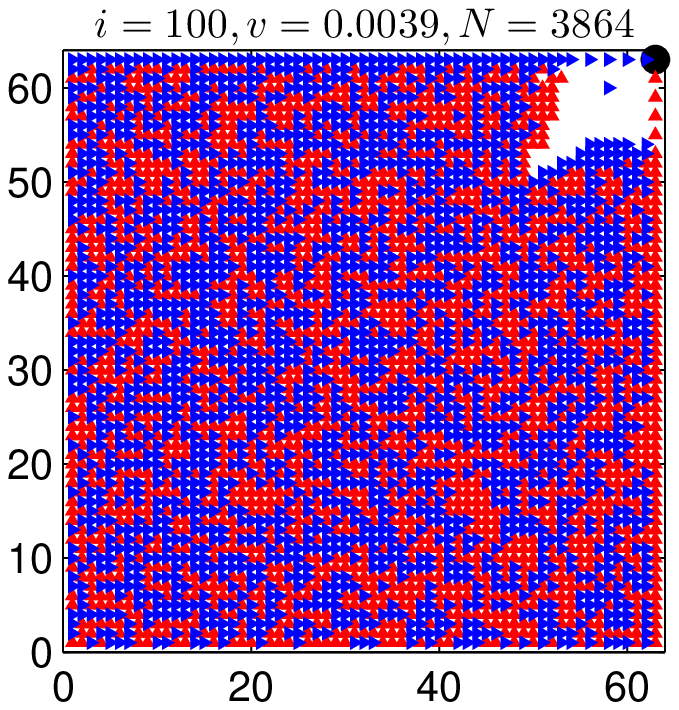}
\end{minipage}}%
\subfigure[]{
\label{o:c}
\begin{minipage}[]{0.33\textwidth}
\centering
\includegraphics[width=1.7in]{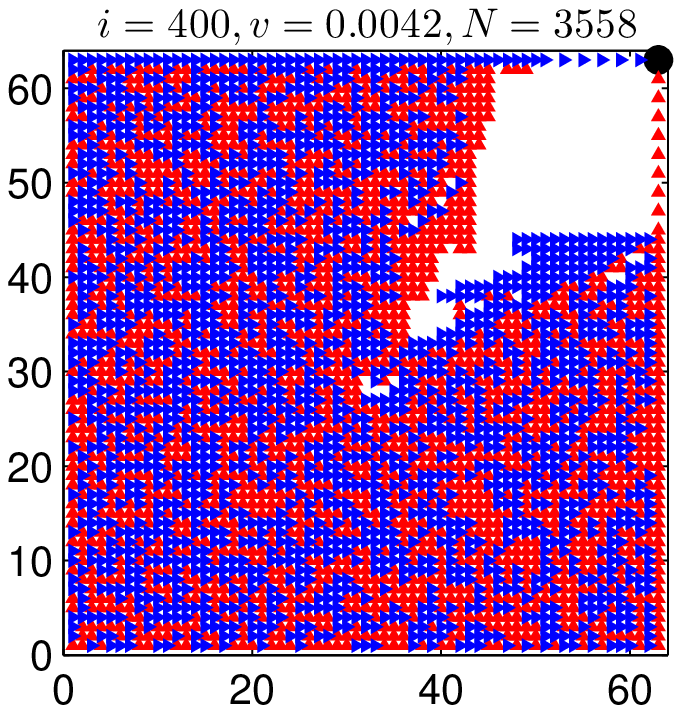}
\end{minipage}} \\[-10pt]
\subfigure[]{
\label{o:d}
\begin{minipage}[]{0.33\textwidth}
\centering
\includegraphics[width=1.7in]{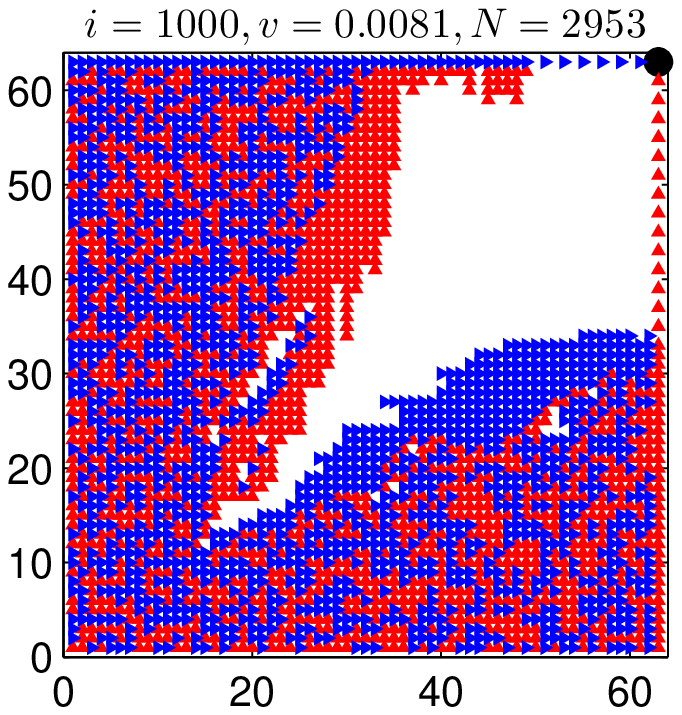}
\end{minipage}}%
\subfigure[]{
\label{o:e}
\begin{minipage}[]{0.33\textwidth}
\centering
\includegraphics[width=1.7in]{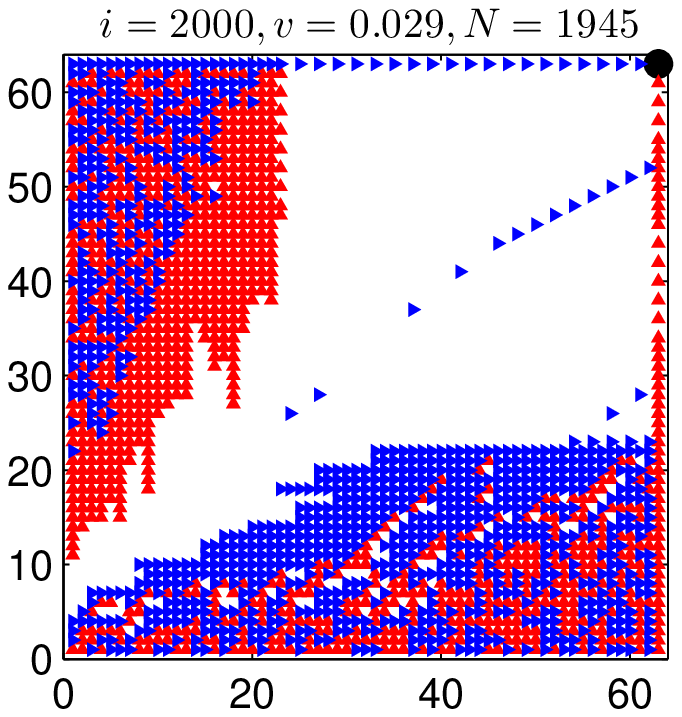}
\end{minipage}}%
\subfigure[]{
\label{o:f}
\begin{minipage}[]{0.33\textwidth}
\centering
\includegraphics[width=1.7in]{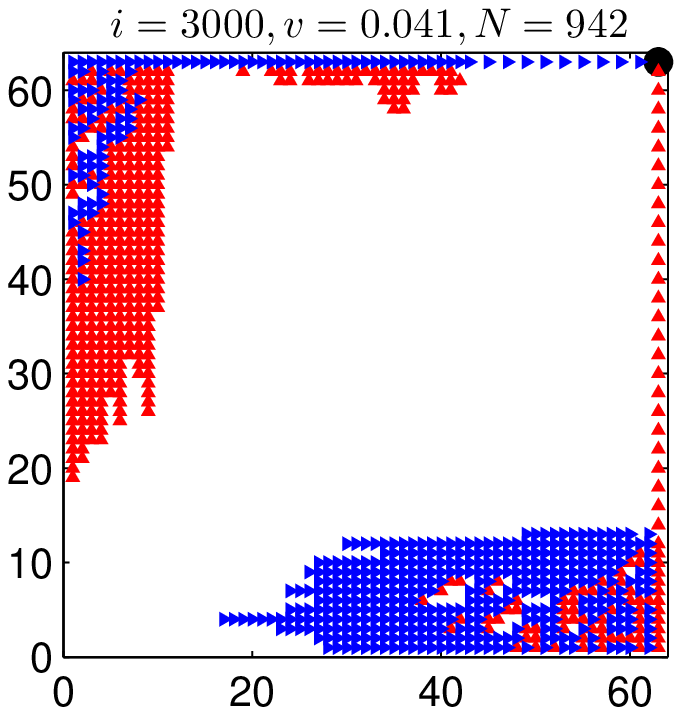}
\end{minipage}}
\caption{The typical configurations observed at several stages of a certain implement in the city of size $L=64$ with a single workplace:(a) in original state; (b) after 100 iterations; (c) after 400 iterations; (d) after 1000 iterations; (e) after 2000 iterations; (f) after 3000 iterations. The black point at the top right corner of each subfigure stands for the workplace (See the black points (64,64). This is just for the convenience of demonstration. For the periodic boundary used in our model, the position of the workplace has no effect on simulation results). On the top of each subfigure, $i$, $v$, $N$ stand for the current number of iteration, current average velocity, cars leaving in the city, respectively. }
\label{o}
\end{figure}

In this section, we will discuss the Scenario-4, i.e. the workplace narrows down to include one site, which is the extreme case of Scenario-3. We assume that this site, which acts as a ``black hole", can accommodate infinite number of cars. We mainly investigate the velocity diagrams and the typical jam configurations. In Fig.~\ref{ov}, we plot the velocity diagrams for the single workplace and double workplaces each of which has only one site.

For the city with a single workplace, all cars can arrive at their destinations and never get into completely traffic jam. If the workplace is in one piece including more than one site, cars going for different goals may be obstructed each other and the gridlock will form inside the workplace. But when the workplace includes one site, this inner obstruction will not happen and all cars will fall into the ``black hole" even if the city is full of cars ($\rho=1$). Fig.~\ref{o} shows the typical configurations observed at several stages of a certain implement in the city of size $L=64$. Initially, all lattice sites are occupied by cars except the workplace. All cars get ready to go to the workplace (that black point). The empty area spreads out over time from top right corner to lower left corner. The center of the empty area is almost square, which is due to the way of route assignment in our model (See Fig.~\ref{route}).

\begin{figure}[htb]
\centerline{\psfig{file=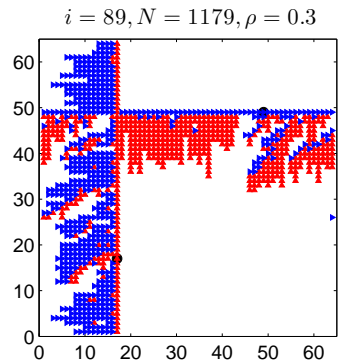,width=7cm}}
\caption{The typical jam configuration observed in a certain implement in the city of size $L=64$ with double workplaces, each of which has only one site. (See the black points (17,17) and (49,49)). On the top of the figure, $i$, $N$, $\rho$ stand for the current number of iteration, cars leaving in the city, initial density of cars respectively. }
\label{o2}
\end{figure}

For the city with double workplaces each of which has one site, the frozen-jam will happen quickly even though the density is very low. In our simulation, we randomly assign one of the two sites to half of cars as their destinations. Each workplace site can't hold the other half of cars which don't belong to it. Since their starting points are randomly distributed, cars leaving for different goals will be obstructed each other and go into complete jams soon. Fig.~\ref{o2} shows the typical jam configuration observed in a certain implement in the city of size $L=64$.

\subsubsection{The finite-size effects of our model}
\begin{figure}[htb]
\subfigure[]{
\label{cd:a} %% label for first subfigure
\begin{minipage}[b]{0.5\textwidth}
\centering
\includegraphics[width=2.6in]{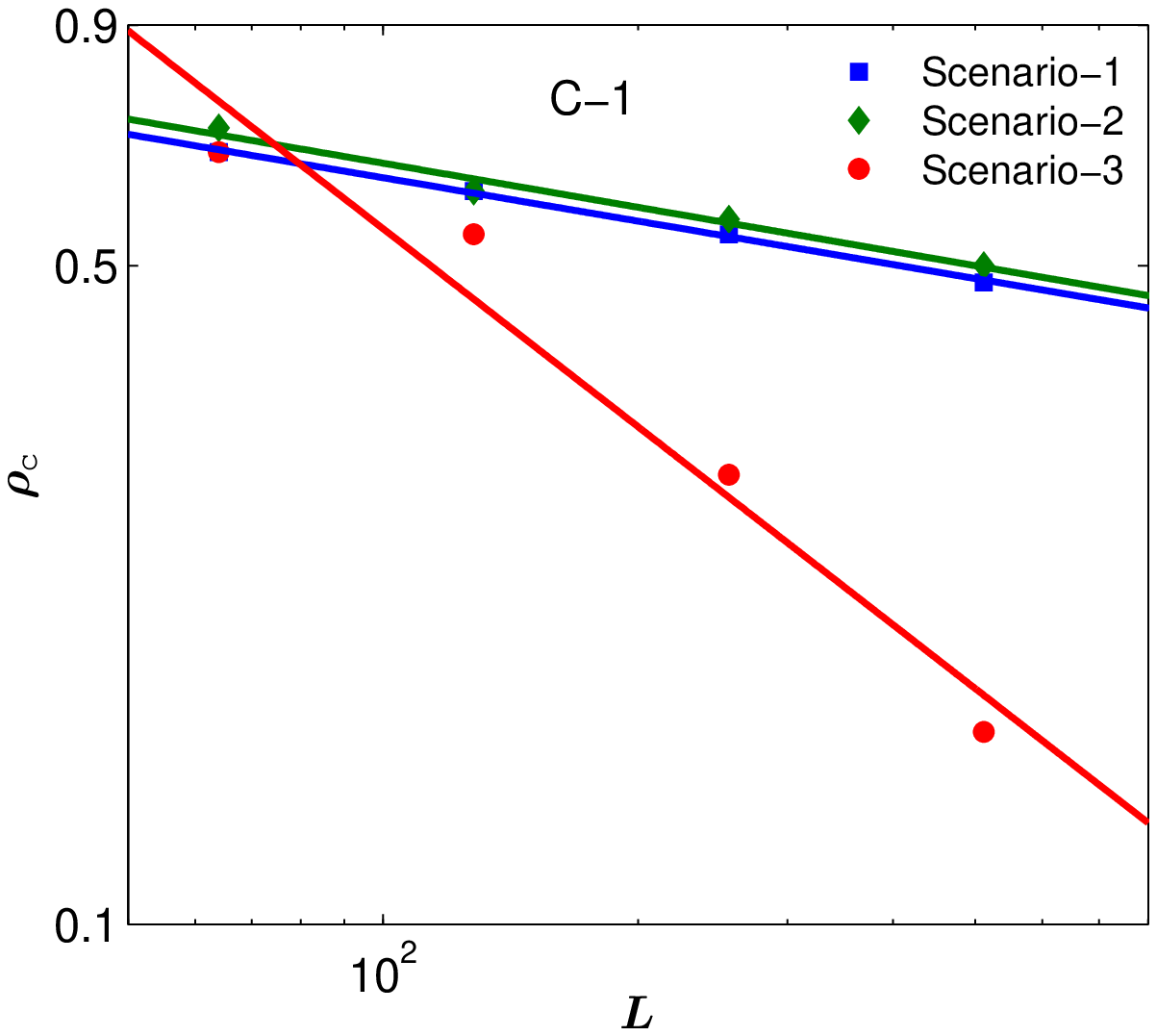}
\end{minipage}}%
\subfigure[]{
\label{cd:b} %% label for second subfigure
\begin{minipage}[b]{0.5\textwidth}
\centering
\includegraphics[width=2.6in]{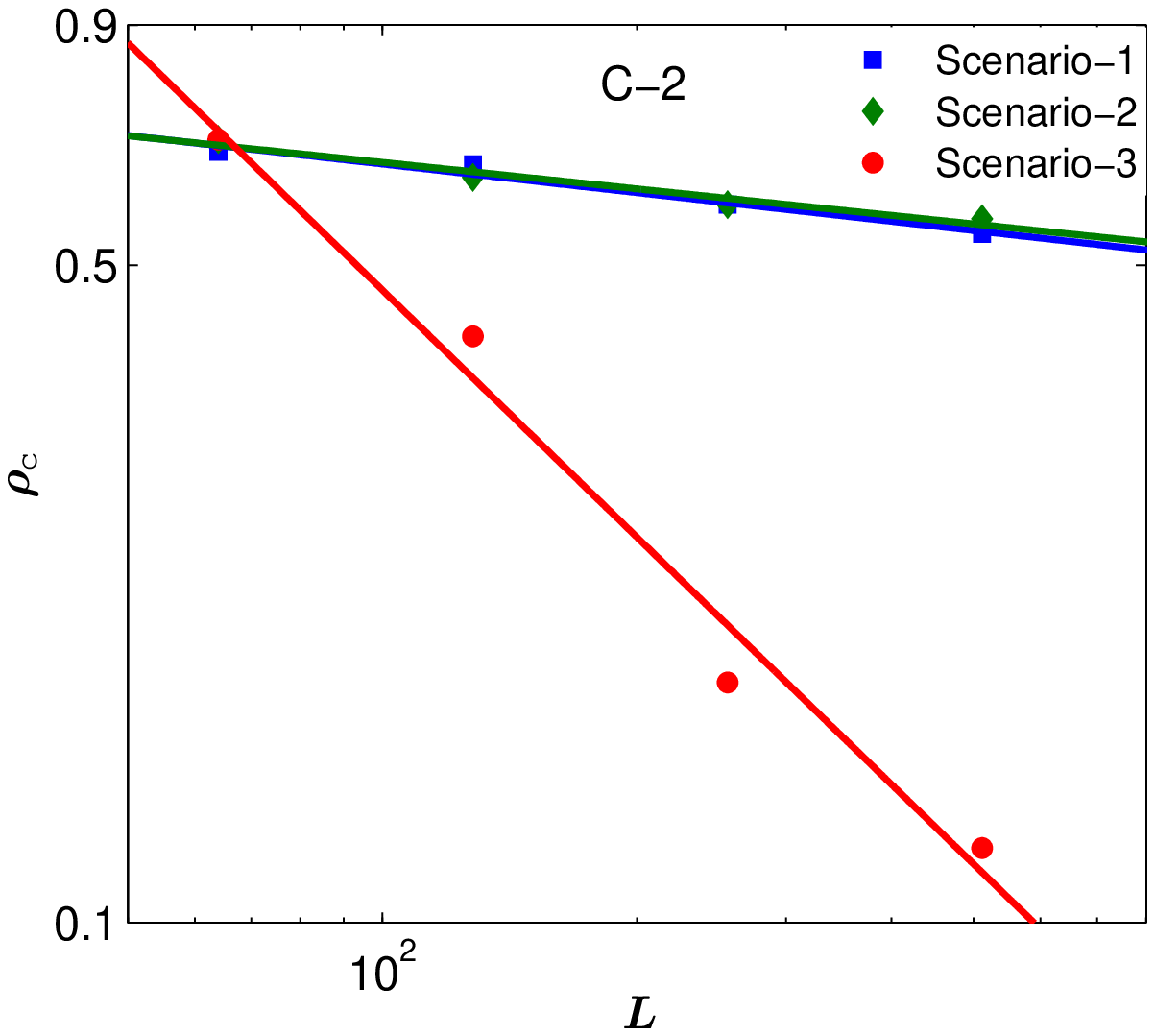}
\end{minipage}}
\caption{Log-log plot of $\rho_{c}(L)$ with the lattice size $L=64,128,256,512$ in Scenario-1,2,3. (a) the city with a single workplace (C-1); (b) the city with double workplaces (C-2).
}
\label{cd} %% label for entire figure
\end{figure}

The average velocity shown in Figs.~\ref{vr},\ref{ov} shows strong finite-size effects, similar to what is known for other 2d traffic models.\cite{Biham1992, Torok1996, Benyoussef2003, Shi2007, In-nami2007} As defined in subsection \ref{vd}, the critical density ($\rho_{c}$) is the density above which typical jamming configurations form, and thus ensemble average velocity is nearly 0. The relationship between the critical density and lattice size is able to be depicted by the exponential model in Ref.~\refcite{Shi1999}, that is,
\begin{equation}
\rho_{c}(L)=\beta L^{\alpha}
\label{model}
\end{equation}
where $\beta$ is a coefficient, while $\alpha$ is the exponent. We use our simulation results to verify this hypothesis and obtain the approximate value of $\alpha$ and $\beta$ with the linear regression method, which are shown in Fig.~\ref{cd} and Table.~\ref{tab}. In order to conduct the goodness-of-fit testing, the sample variance ($s^{2}$) is adopted, i.e. $s^{2}=Q/(n-2)$, where $Q$ is the residual sum of squares (RSS) and $n$ is the sample size. The sample variance is an unbiased estimation of fit error variance.

\begin{table}[htb]
\tbl{Estimation of model parameters and fit error variance}
{\begin{tabular}{@{}ccccc@{}} \toprule
Scenario & $\beta_{1}$, $\alpha_{1}$ & $\beta_{2}$, $\alpha_{2}$ & $s^{2}_{1}$ & $s^{2}_{2}$ \\ \colrule
1 & \hphantom{0}1.25, -0.15 & \hphantom{0}1.02, -0.10 & 0.000063 & 0.00043 \\
2 & \hphantom{0}1.31, -0.16 & \hphantom{0}0.99, -0.09 & 0.00063\hphantom{0} & 0.00041 \\
3 & 13.63, -0.70 & 26.21, -0.87 & 0.026\hphantom{000} & 0.017\hphantom{00} \\
4 & --- & \hphantom{0}2.77, -0.65 & ---\ & 0.048\hphantom{00} \\ \botrule
\end{tabular}
\label{tab}}
\end{table}

In Table.~\ref{tab}, $\beta_{1}$ and $\alpha_{1}$ are the estimation of the coefficient and exponent of model \ref{model} with the values obtained from C-1. Similarly, $\beta_{2}$, $\alpha_{2}$ are the estimation from C-2. $s^{2}_{1}$ and $s^{2}_{2}$ are the sample variance corresponding to C-1 and C-2 respectively. In particular, in Scenario-4 of C-1 the complete traffic jam will never happen even if the city is full of cars (see subsection \ref{sce-4} or Fig.~\ref{ov:a}). So in that case the critical density dose not exist, which is denoted by dashes (--) in Table.~\ref{tab}.

As shown in Fig.~\ref{cd} and Table.~\ref{tab}, we obtain a good fit to Eq.~(\ref{model}) in the Scenario-1,2 whether for C-1 or C-2. But a major error will occur when it is applied to Scenario-3,4. From the simulation data available now, it is unclear whether the exponential model (\ref{model}) will work, i.e. the critical density and lattice size will remain  the linear relationship approximately in the logarithmic coordinates, when the area ratio ($r$) of the workplace to the city in Scenario-1,2 increases higher than 20\%. This is worth carrying out increasingly comprehensive simulations and further research.

\subsection{Arrival time probability distributions}

\begin{figure}[htb]
\subfigure[]{
\label{A1:a} %% label for first subfigure
\begin{minipage}[b]{0.5\textwidth}
\centering
\includegraphics[width=2.6in]{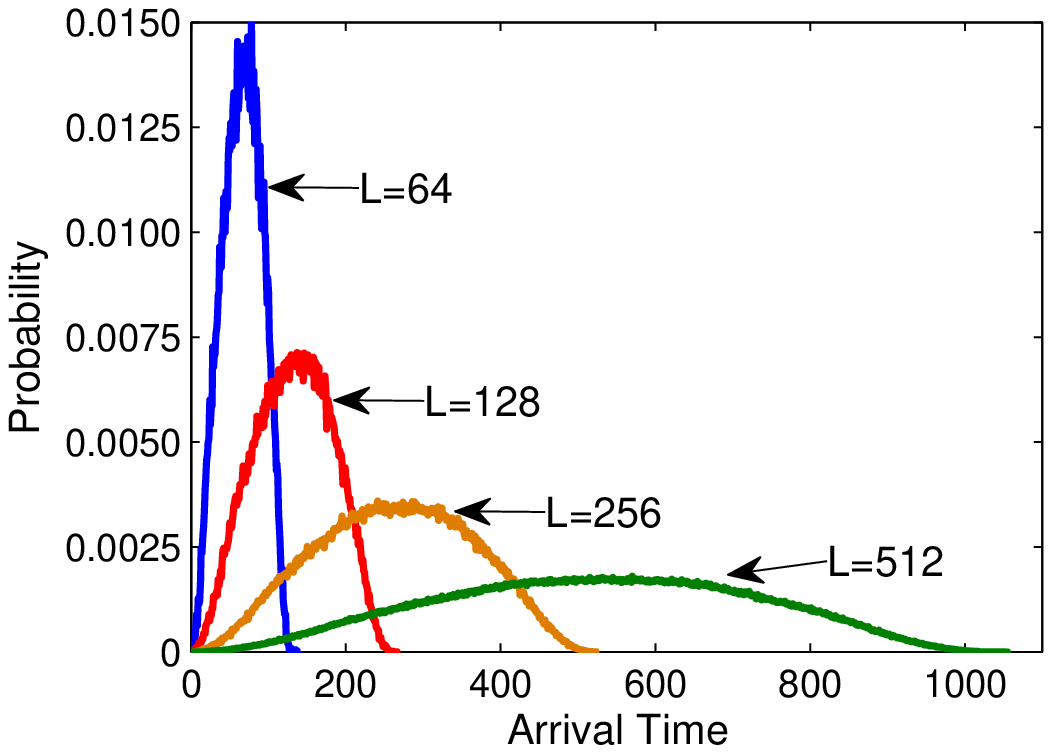}
\end{minipage}}%
\subfigure[]{
\label{A1:b} %% label for second subfigure
\begin{minipage}[b]{0.5\textwidth}
\centering
\includegraphics[width=2.6in]{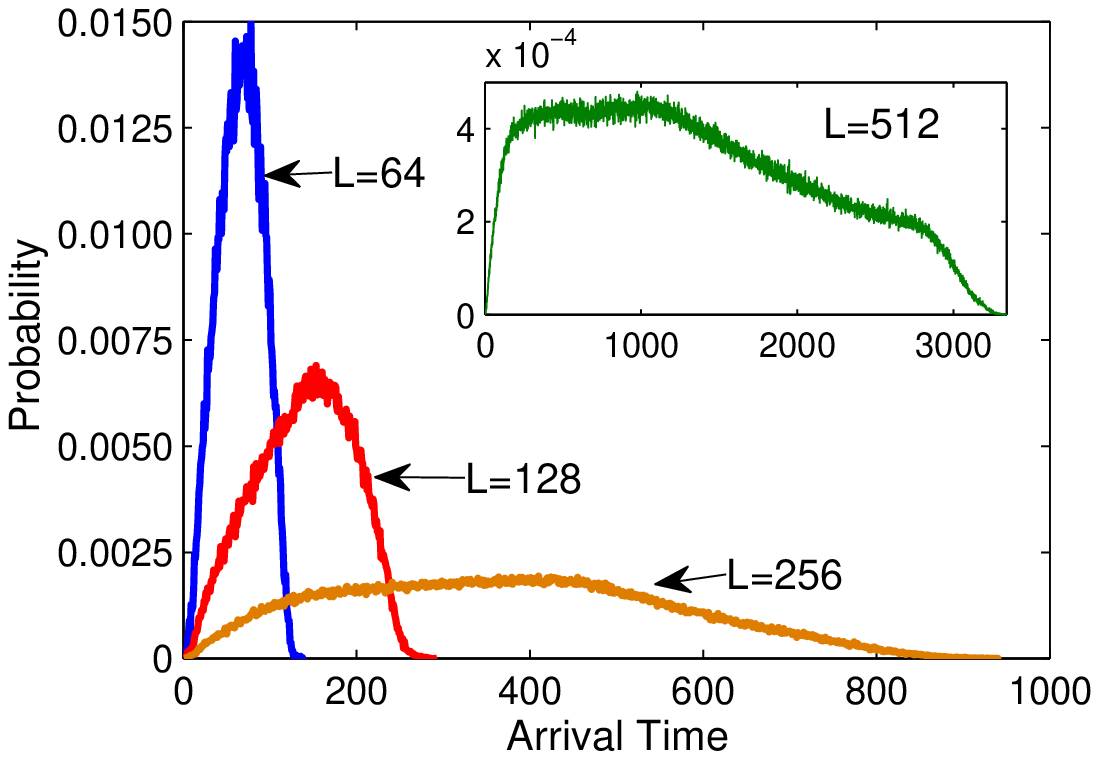}
\end{minipage}}
\caption{Effect of different city sizes on the arrival time probability distributions at relatively low density ($\rho=0.1$) in the city with a single workplace in (a) the Scenario-1; (b) the Scenario-3. The curve  are averaged out by one hundred samples. Because the range of probability of size $L=512$ is one order of magnitude lower than the size $L=256$, its diagram is plotted especially as a inset within Fig.~\ref{A1:b} for a clearly presentation. These diagrams are plotted in a histogram form with the bin granularity of one time step.
}
\label{A1} %% label for entire figure
\end{figure}

The effect of city sizes on the arrival time probability distribution (ATPD) at relatively low density are shown in Fig.~\ref{A1}. We compare the simulation results in the Scenario-1 with Scenario-3 considering four city sizes. The distributions of all sizes in Fig.~\ref{A1:a} are unimodal and symmetrical. The position of every curve peak equals approximately its corresponding city size. However, the curves shapes of large systems in Fig.~\ref{A1:b} are asymmetrical with a flat front part (e.g. $L=256$ and $L=512$) and the maximal arrival times are far greater than that in Fig.~\ref{A1:a} for large systems.

\begin{figure}[htb]
\subfigure[]{
\label{A_ruo:a} %% label for first subfigure
\begin{minipage}[b]{0.5\textwidth}
\centering
\includegraphics[width=2.5in]{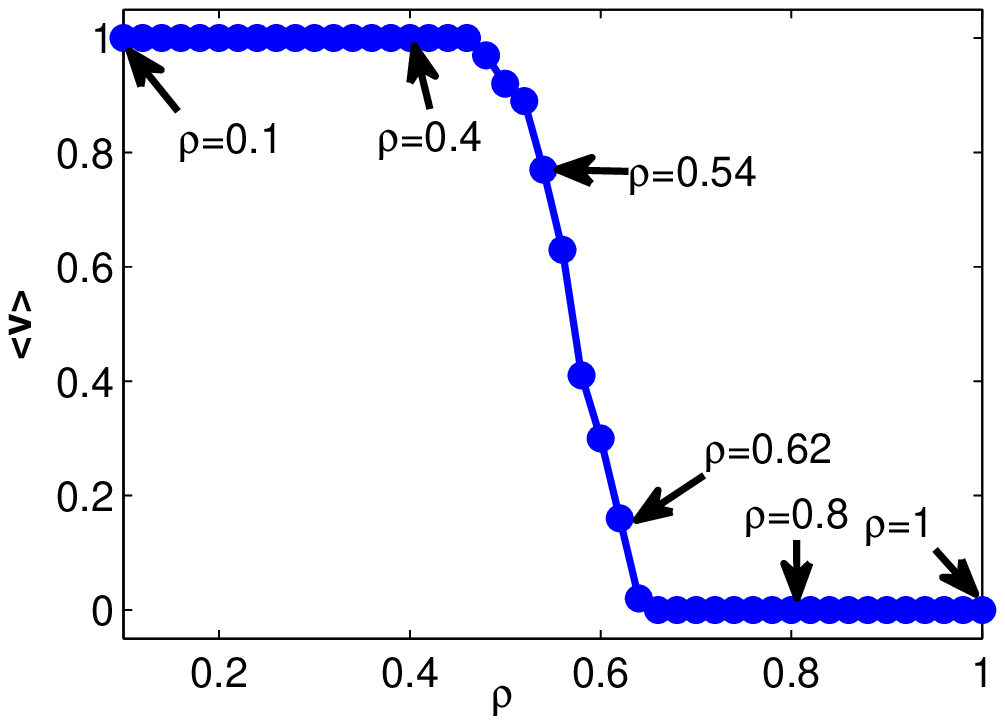}
\end{minipage}}%
\subfigure[]{
\label{A_ruo:b} %% label for second subfigure
\begin{minipage}[b]{0.5\textwidth}
\centering
\includegraphics[width=2.5in]{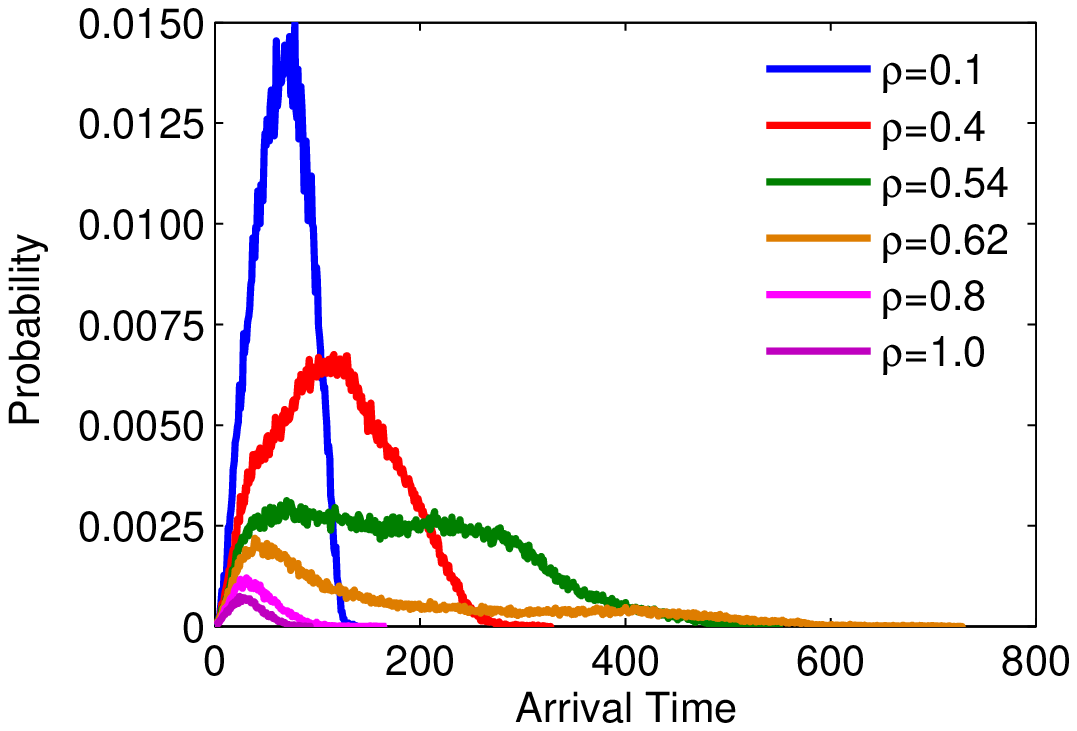}
\end{minipage}}
\caption{The arrival time probability distribution in different initial densities in the city with a single workplace ($L=64$, $M=20$). (a) Six typical densities $\rho=0.1, 0.4, 0.54, 0.62, 0.8, 1.0$ are selected to be surveyed. (b) The ATPD curves corresponding the six densities. The curves are averaged out by one hundred samples at every surveyed density. When the traffic gets into a complete jam in one realization, the arrival time steps of cars leaving in the city are considered to be infinite and not concluded in the statistical data. The diagram (b) is plotted in a histogram form with the bin granularity of one time step.
}
\label{A_ruo} %% label for entire figure
\end{figure}

%\begin{table}[]
%\tbl{The statistics on simulation data in Fig.~\ref{A_ruo}}
%{\begin{tabular}{@{}ccccc@{}} \toprule
%Letter & Density & Maximal & Maximal & Arrival \\
%symbol & values & arrival time & probability & rate (\%) \\ \colrule
%A & 0.1\hphantom{0} & 78 & 0.015\hphantom{00} & 1.0\hphantom{00} \\
%B & 0.4\hphantom{0} & 116 & 0.0068\hphantom{0} & 1.0\hphantom{00} \\
%C & 0.54 & 70 & 0.0031\hphantom{0} & 0.85\hphantom{0} \\
%D & 0.62 & 39 & 0.0022\hphantom{0} & 0.34\hphantom{0} \\
%E & 0.8\hphantom{0} & 31 & 0.0012\hphantom{0} & 0.055 \\
%F & 1.0\hphantom{0} & 24 & 0.00074 & 0.027 \\ \botrule
%\end{tabular} \label{ta1}}
%\end{table}

The arrival time probability distributions in different initial densities in the city with a single workplace is shown in Fig.~\ref{A_ruo}. We find that with the increase of density, the arrival time of maximal probability increases firstly, and then decreases since the density reaches the transition region in the velocity diagram. In the low density region of free moving phase, all cars can arrive at their destinations in a relatively short time (see the curves corresponding A, B in Fig.~\ref{A_ruo}). When the density reaches the transition region (see the curves corresponding C, D in Fig.~\ref{A_ruo}), only a part of cars can arrive at their destinations and their arrival time steps are extended. When the density reaches the region of jam phase, only a small percentage of cars can arrive at their destinations. Even within dozens of time steps, the rest cars self-organize into one large cluster and can not go ahead any more (see the curves corresponding E, F in Fig.~\ref{A_ruo}). The number of arrived cars equals $N\cdot\int_{0}^{T_{max}}p(t)dt$, i.e. the area under the curve multiplied by the total number of cars, where $p$ is the arrival time probability, $T_{max}$ is the maximal arrival time. Considering the discrete-valued arrival time, the expression should be transformed to $N\cdot\sum_{i=1}^{T_{max}}p_{i}$, where $p_{i}$ is the arrival time probability of the $i$-th time step.

\begin{figure}[htb]
\subfigure[]{
\label{A3:a}
\begin{minipage}[]{0.5\textwidth}
\centering
\includegraphics[width=2.6in]{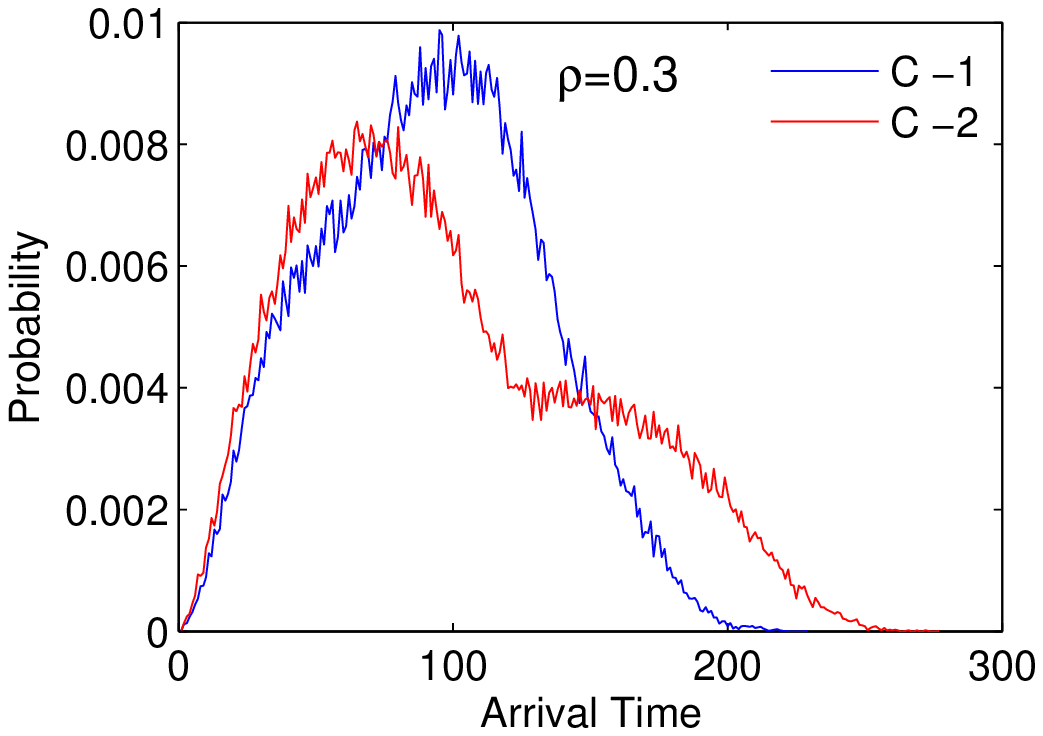}
\end{minipage}}%
\subfigure[]{
\label{A3:b}
\begin{minipage}[]{0.5\textwidth}
\centering
\includegraphics[width=2.6in]{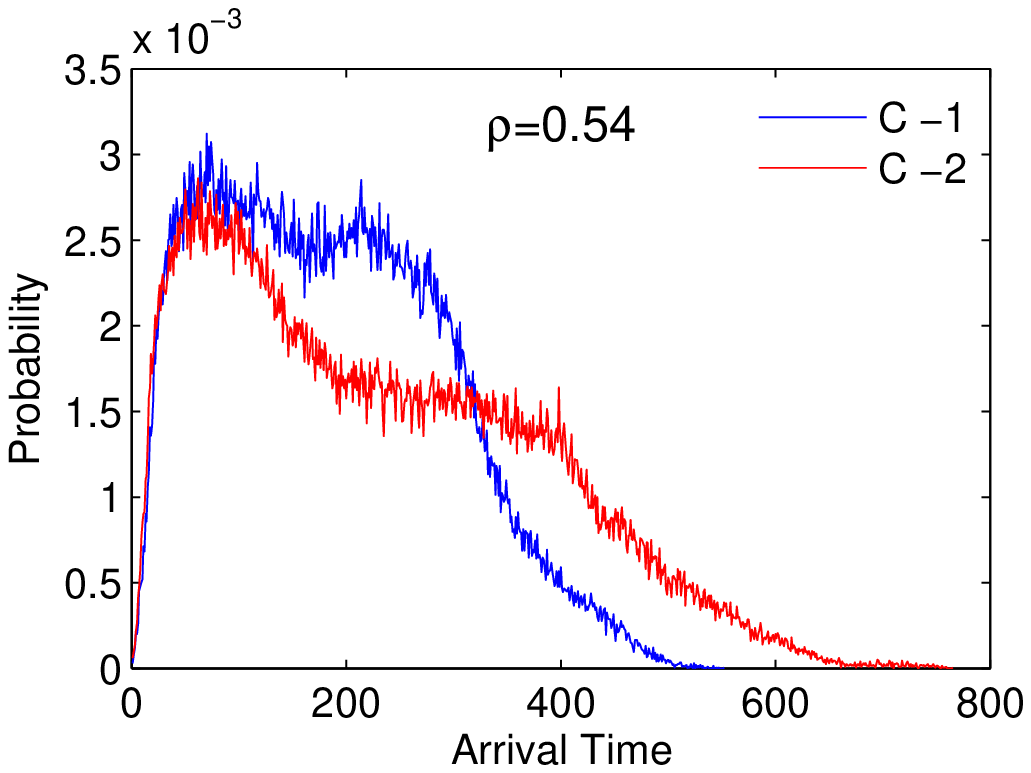}
\end{minipage}} \\
\subfigure[]{
\label{A3:c}
\begin{minipage}[]{0.5\textwidth}
\centering
\includegraphics[width=2.6in]{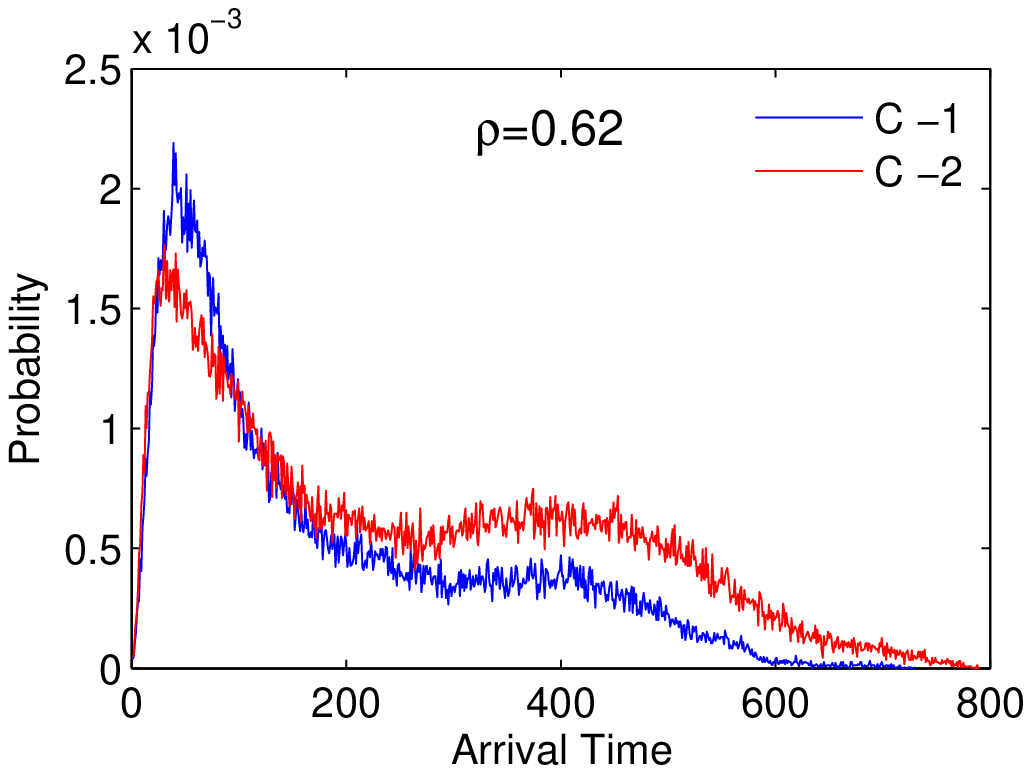}
\end{minipage}}%
\subfigure[]{
\label{A3:d}
\begin{minipage}[]{0.5\textwidth}
\centering
\includegraphics[width=2.6in]{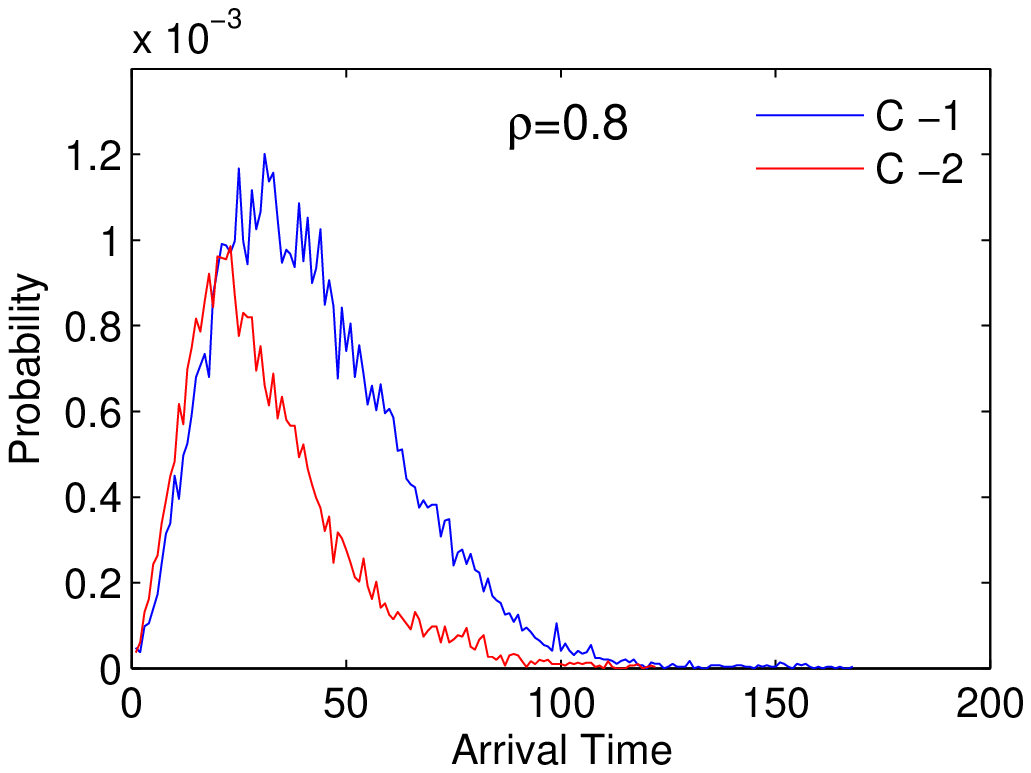}
\end{minipage}}
\caption{The arrival time probability distribution at the city with a single workplace ($L=64$, $M=20$) is compared with double workplaces ($L=64$, $M=14.14$). Four typical densities are selected to be surveyed corresponding (a) $\rho=0.3$; (b) $\rho=0.54$; (c) $\rho=0.62$; (d) $\rho=0.8$. The curves are averaged out by one hundred samples at every surveyed density. When the traffic gets into a complete jam in one realization, the arrival time steps of cars left over in the city are considered to be infinite and not included in the statistical data. These diagrams are plotted in a histogram form with the bin granularity of one time step.
}
\label{A3}
\end{figure}

Fig.~\ref{A3} shows the arrival time probability distributions of C-1 and C-2 in four typical densities. We observe that the peak positions of C-2 move to left a tiny bit with respect to C-1. But the maximal arrival times of C-2 are larger than those of C-1 except for $\rho=0.8$ in Fig.~\ref{A3:d}. Comparing the area under probability curves of C-1 with C-2, we find that the number of arrived cars of C-1 are almost the same as C-2 in the low density region (see Fig.~\ref{A3:a} and Fig.~\ref{A3:b}). The number of arrived cars are greater for C-2 than C-1 in the intermediate density region (see Fig.~\ref{A3:c}), but less in the high density region (see Fig.~\ref{A3:d}). Fig.~\ref{A3:d} indicates that the traffic will be trapped easily and quickly for C-2 than C-1 in the region of high density.

The travel time can't be saved in C-2 comparing with C-1, even longer, which contradicts the usual experience in reality. We ascribe the experimental phenomena to the model definition which permits only two travel directions, i.e. cars can only move upward or rightward. If cars can move in four directions as do in real road network, the results may be in accordance with actual traffic. The extension of BML model including this character is worth further research.

\subsection{Destination arrival rate}
The destination arrival rate is define as the ratio of cars arriving destinations before the traffic gets into a complete jam to all cars in the city. If all cars have arrived at their terminals finally, the arrival rate is set to one.
\begin{figure}[htb]
\subfigure[]{
\label{d:a}
\begin{minipage}[]{0.33\textwidth}
\centering
\includegraphics[width=1.8in]{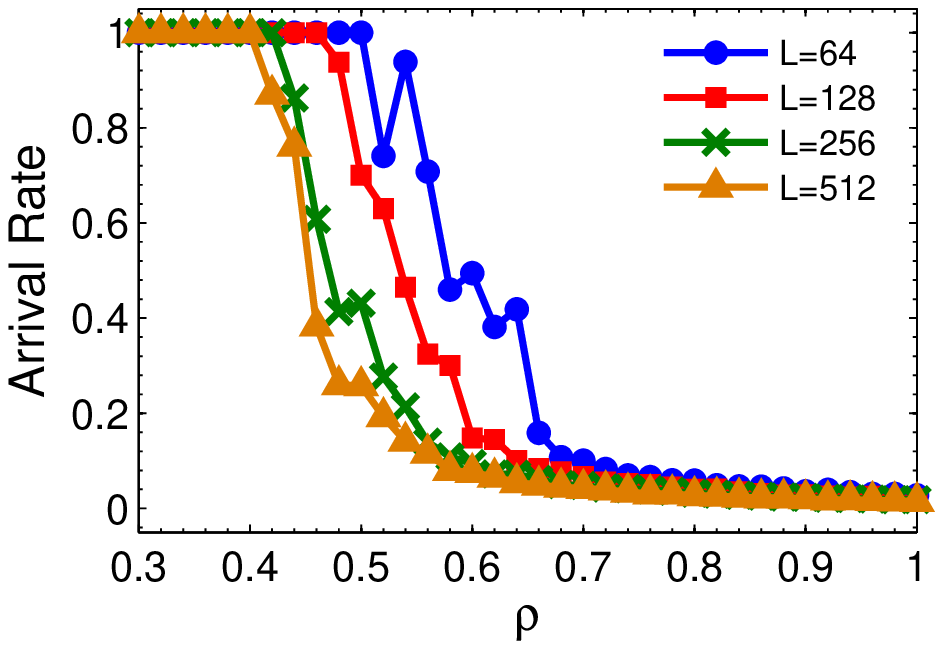}
\end{minipage}}%
\subfigure[]{
\label{d:b}
\begin{minipage}[]{0.33\textwidth}
\centering
\includegraphics[width=1.8in]{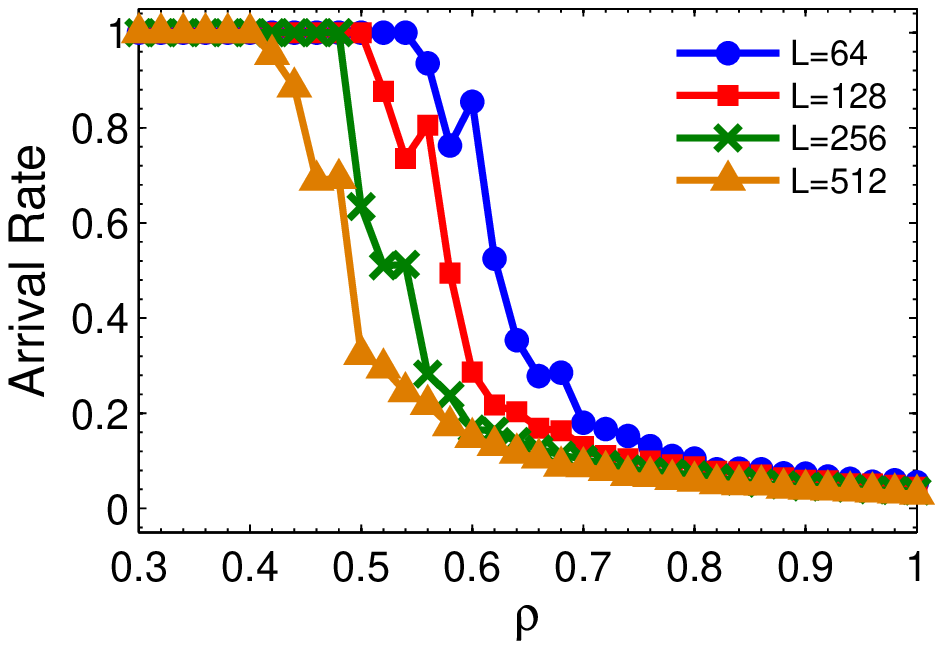}
\end{minipage}}%
\subfigure[]{
\label{d:c}
\begin{minipage}[]{0.33\textwidth}
\centering
\includegraphics[width=1.8in]{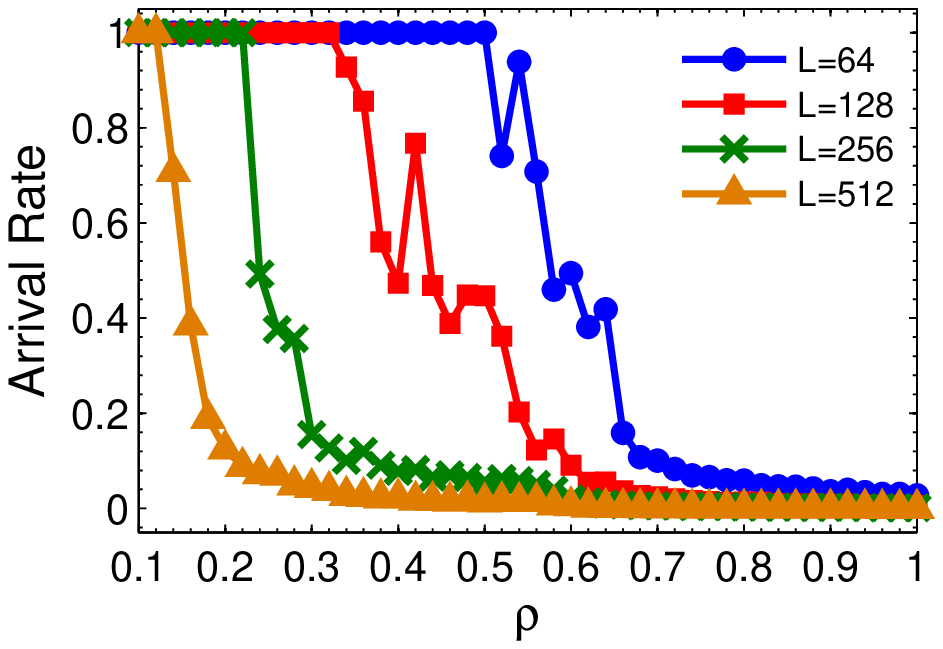}
\end{minipage}} \\
\subfigure[]{
\label{d:d}
\begin{minipage}[]{0.33\textwidth}
\centering
\includegraphics[width=1.8in]{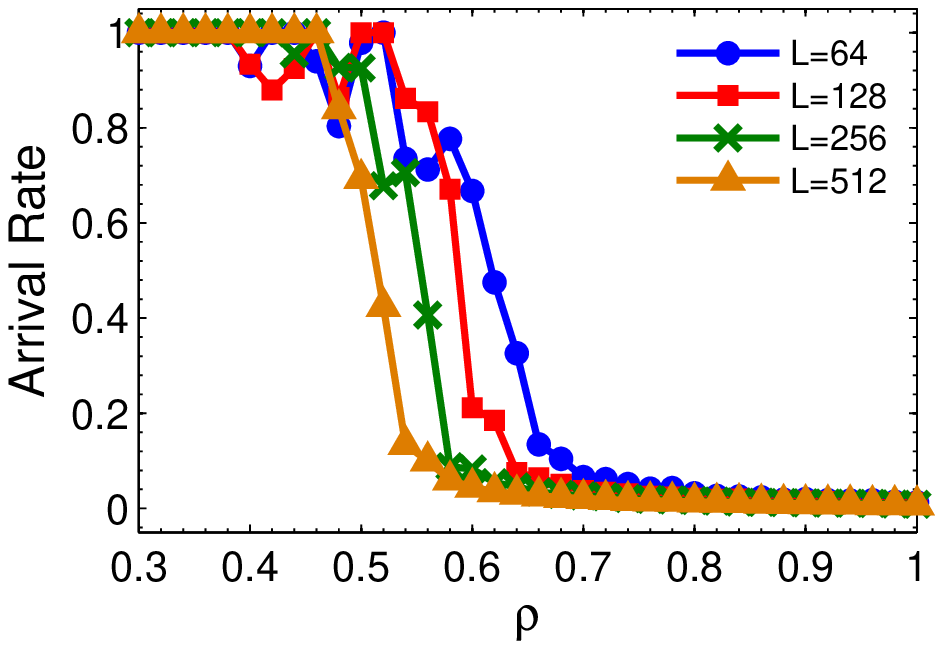}
\end{minipage}}%
\subfigure[]{
\label{d:e}
\begin{minipage}[]{0.33\textwidth}
\centering
\includegraphics[width=1.8in]{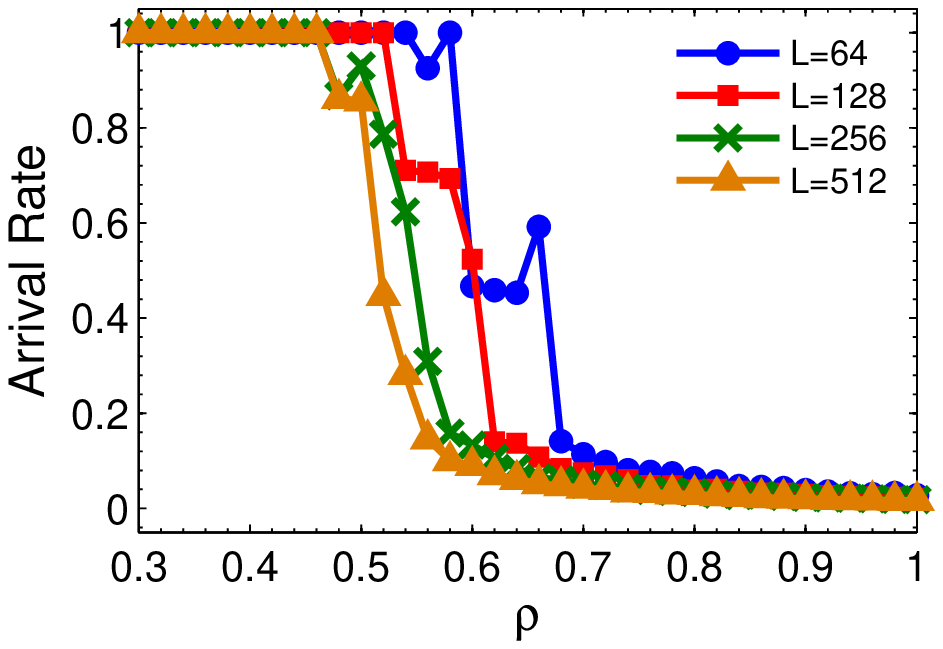}
\end{minipage}}%
\subfigure[]{
\label{d:f}
\begin{minipage}[]{0.33\textwidth}
\centering
\includegraphics[width=1.8in]{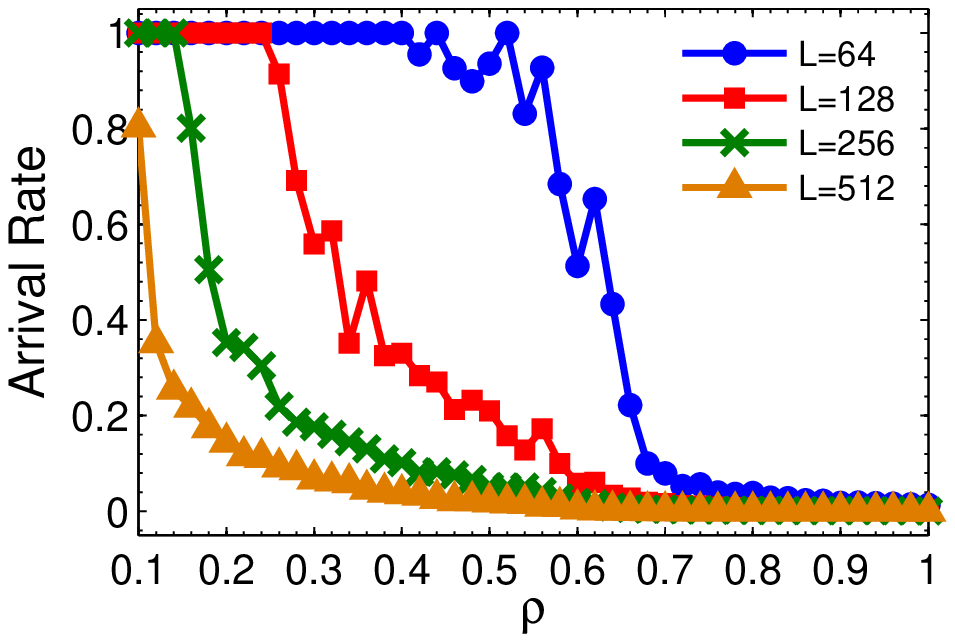}
\end{minipage}}
\caption{The destination arrival rate versus density ($\rho$) in the city of (a) C-1 in the Scenario-1; (b) C-1 in the Scenario-2; (c) C-1 in the Scenario-3; (d) C-2 in the Scenario-1; (e) C-2 in the Scenario-2; (f) C-2 in the Scenario-3. The surveyed density increases with the step length of $0.02$. The destination arrival rate in every density is averaged out by ten implements.
}
\label{d}
\end{figure}

Fig.~\ref{d} shows the destination arrival rate as a function of the density for the four different city sizes and two types of urban layouts. The trend derived from these six diagrams is very similar to the velocity diagrams shown in Fig.~\ref{vr}. As the city size enlarges, the curve moves to left and becomes smoother. The traffic in the Scenario-3 is worse than Scenario-1,2. It is important to note that the arrival rate of C-2 falls faster than C-1 in the transition region except for in the Scenario-3.

\subsection{Convergence time}

\begin{figure}[htb]
\subfigure[]{
\label{X:a}
\begin{minipage}[]{0.25\textwidth}
\centering
\includegraphics[width=1.3in]{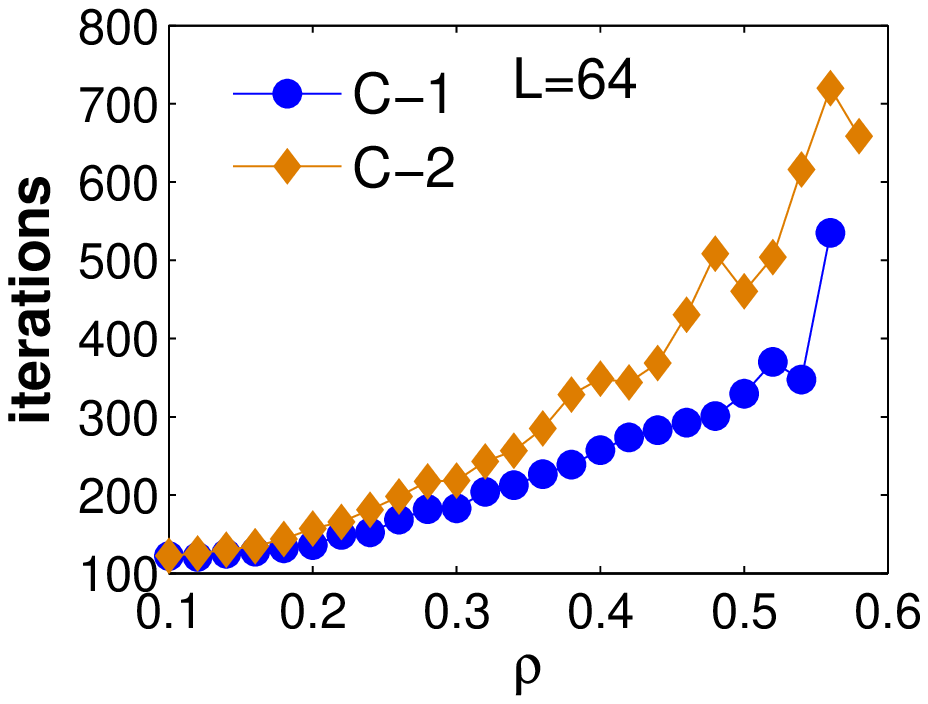}
\end{minipage}}%
\subfigure[]{
\label{X:b}
\begin{minipage}[]{0.25\textwidth}
\centering
\includegraphics[width=1.3in]{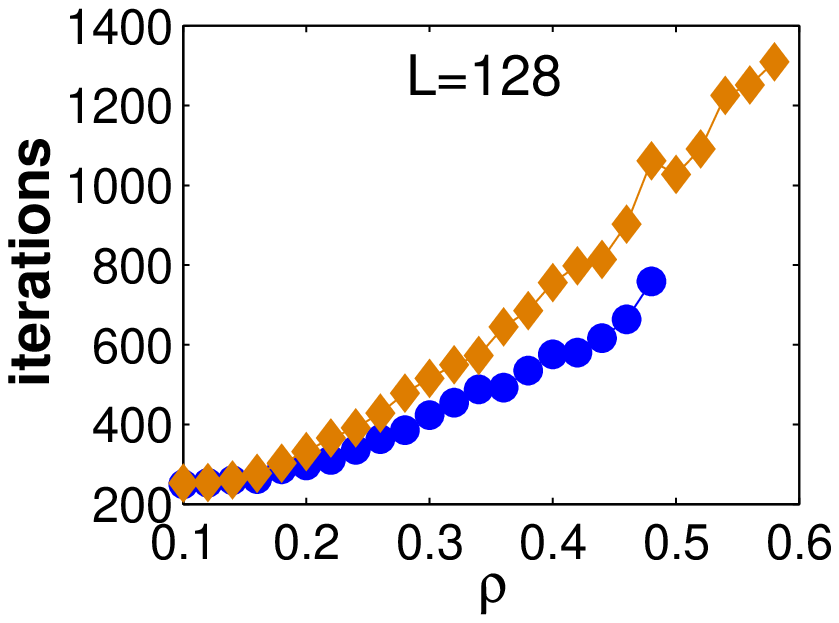}
\end{minipage}}%
\subfigure[]{
\label{X:c}
\begin{minipage}[]{0.25\textwidth}
\centering
\includegraphics[width=1.3in]{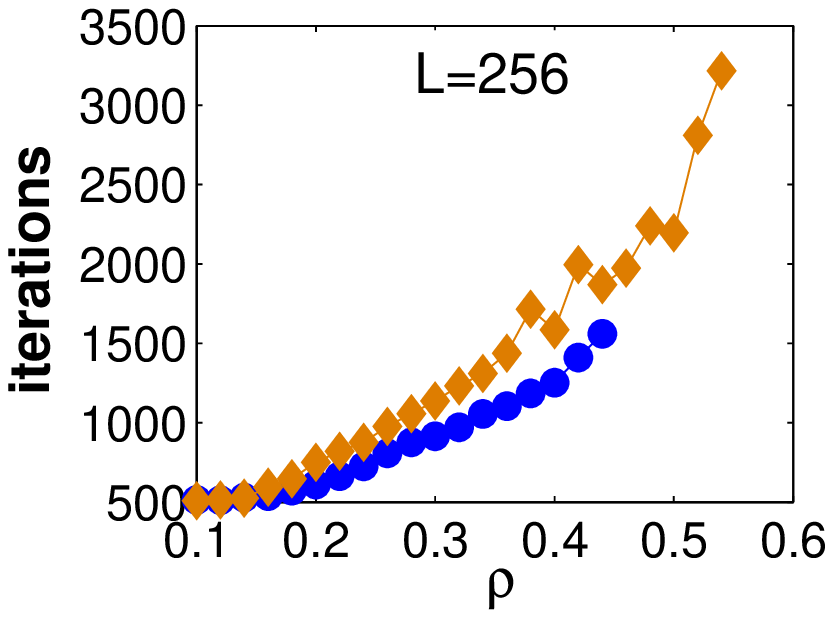}
\end{minipage}}%
\subfigure[]{
\label{X:d}
\begin{minipage}[]{0.25\textwidth}
\centering
\includegraphics[width=1.3in]{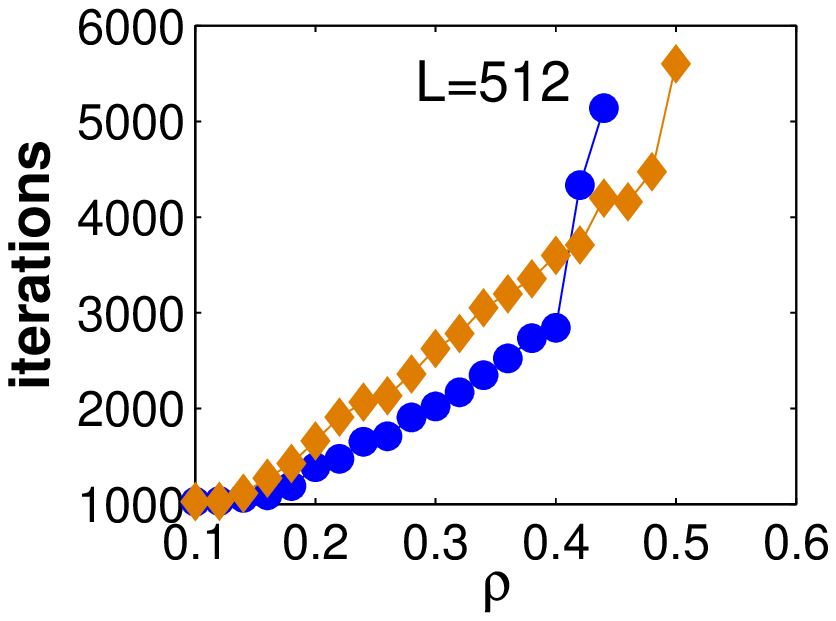}
\end{minipage}}

\subfigure[]{
\label{X:e}
\begin{minipage}[]{0.25\textwidth}
\centering
\includegraphics[width=1.3in]{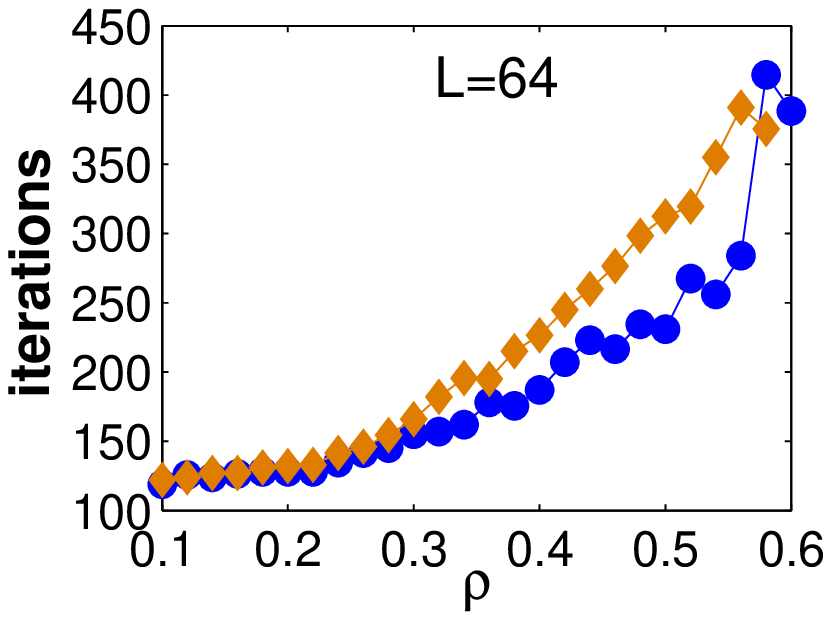}
\end{minipage}}%
\subfigure[]{
\label{X:f}
\begin{minipage}[]{0.25\textwidth}
\centering
\includegraphics[width=1.3in]{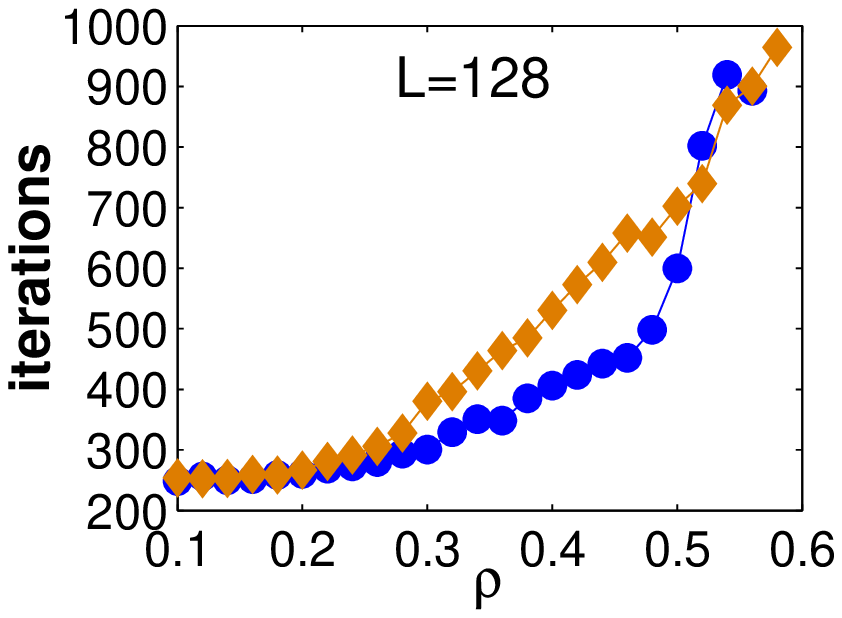}
\end{minipage}}%
\subfigure[]{
\label{X:g}
\begin{minipage}[]{0.25\textwidth}
\centering
\includegraphics[width=1.3in]{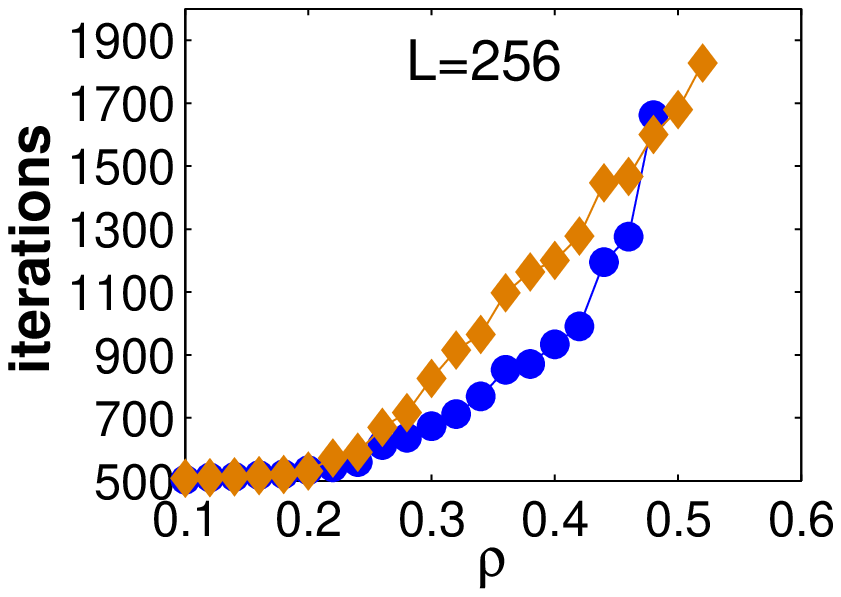}
\end{minipage}}%
\subfigure[]{
\label{X:h}
\begin{minipage}[]{0.25\textwidth}
\centering
\includegraphics[width=1.3in]{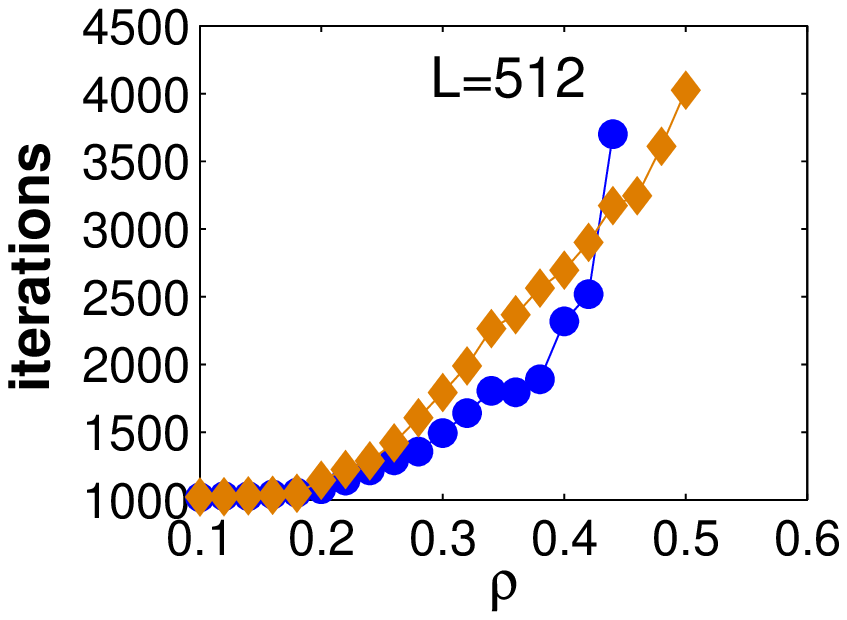}
\end{minipage}}

\subfigure[]{
\label{X:i}
\begin{minipage}[]{0.25\textwidth}
\centering
\includegraphics[width=1.3in]{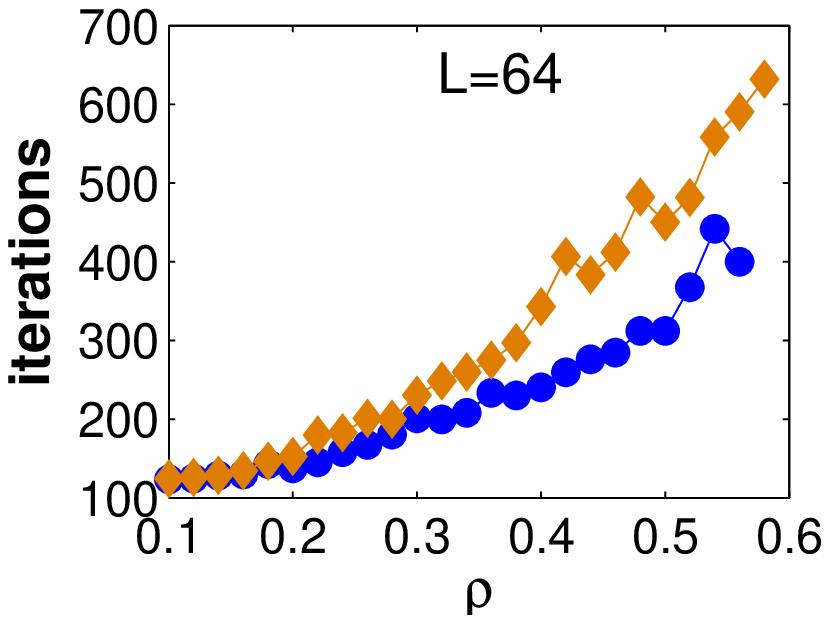}
\end{minipage}}%
\subfigure[]{
\label{X:j}
\begin{minipage}[]{0.25\textwidth}
\centering
\includegraphics[width=1.3in]{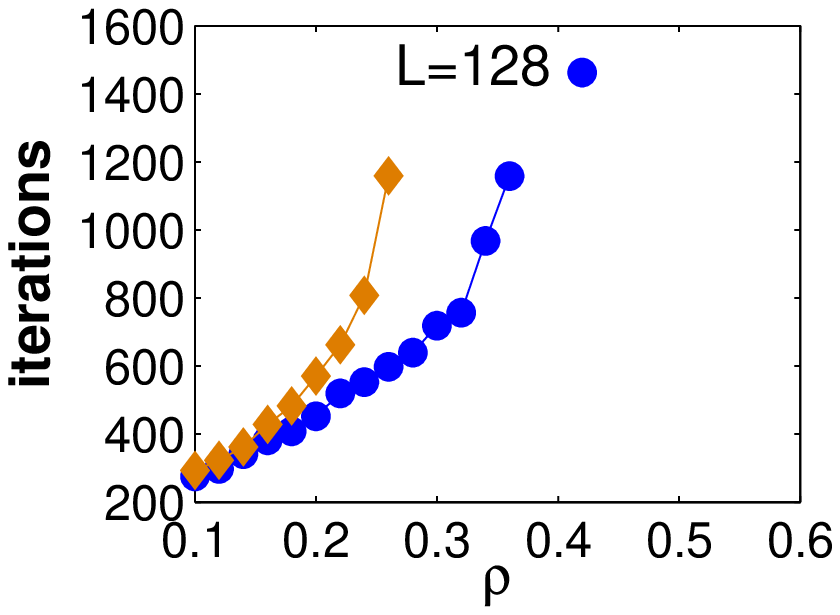}
\end{minipage}}%
\subfigure[]{
\label{X:k}
\begin{minipage}[]{0.25\textwidth}
\centering
\includegraphics[width=1.3in]{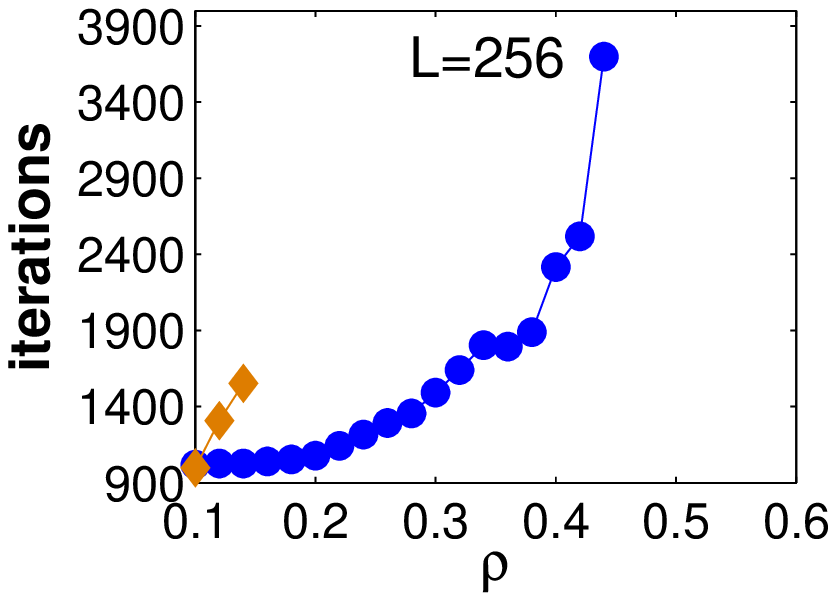}
\end{minipage}}%
\subfigure[]{
\label{X:l}
\begin{minipage}[]{0.25\textwidth}
\centering
\includegraphics[width=1.3in]{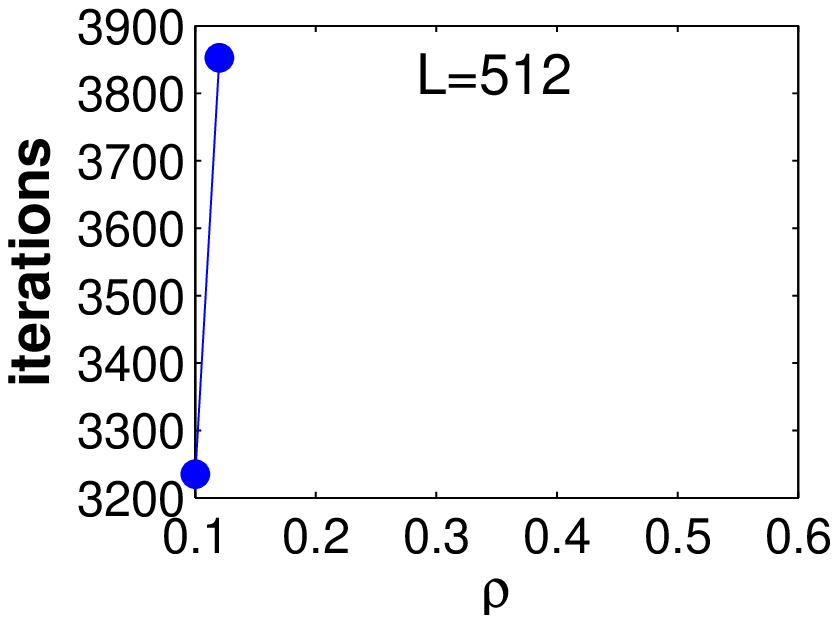}
\end{minipage}}
\caption{Comparison of the convergence time of C-1 with C-2 in three types of scenarios: (a-d) Scenario-1; (e-h) Scenario-2; (j-l) Scenario-3. Ten implements are carried out to find the median value of iteration in every density. If the traffic gets into a complete jam in an implement, the iterations of this implement are set to infinite. When its median value is infinite, the point can not be displayed in the figure (This can explain the discontinuity point in Fig.~\ref{X:j}). In particular, the curve of C-2 is absent in Fig.~\ref{X:l}. This is because the traffic is completely blocked even in the density of $\rho=0.1$.
}
\label{X}
\end{figure}

In this subsection, we measure the convergence time of the simulation by counting the iterations before all cars have arrived at their destinations. We compare the iterations of C-1 with that of C-2 in three types of scenarios and four city sizes in Fig.~\ref{X}. We observe the convergence speed of C-1 is faster than that of C-2 in almost all cases except for the high density region in some instances (see Fig.~\ref{X:d}, \ref{X:e}, \ref{X:f}, \ref{X:h}), which contradicts the usual experience in reality. We ascribe the simulation results to the model definition which permits only two travel directions, just as the interpretation of the simulation in Fig.~\ref{A3}. The number of iteration rises faster in the high density region than the low density region. This is because the interaction between cars becomes strong and car movements become rare as the density increases. So the ratio of time spent waiting at the crossroads rises rapidly with the growth of density.

\section{Conclusions}
\label{conclusion}
In summary, we extend the BML model to describe dynamic route choices between the residence and workplace in cities. We plan to simulate the scenario that office workers drive toward their workplaces from their residences every morning.

We firstly investigate the velocity diagrams considering the four different city sizes ($L=64,128,256,512$) and two types of urban layouts, i.e. the city with a single workplace (C-1) and the city with double workplaces (C-2). The average velocity is greatly influenced by the size of city and the area of workplace. Meanwhile, the urban layout has some effect on the average velocity. We find it would be unwise to enlarge the city while remain the workplace area constant.

We perform a finite-size scaling analysis of the critical density  from a statistical point of view and the order parameter of this jamming transition is estimated. We adopt the exponential model in Ref.~\refcite{Shi1999} to depict the relationship of the critical density and lattice size. The parameters of model are estimated with the method of linear regression. We obtain a good fit to the model in the Scenario-1,2 whether for C-1 or C-2. But a major error will occur when it is applied to Scenario-3,4. From the simulation data available now, it is unclear whether the exponential model (\ref{model}) will work when the area ratio ($r$) in Scenario-1,2 increases higher than 20\%. This is worth carrying out increasingly comprehensive simulations and further research.

The effects of urban size, car density and urban layout on the arrival time probability distributions are also studied. The travel time of drivers can't be saved in C-2 compared with C-1, even longer, which contradicts the usual experience in reality. We ascribe the simulation results to the model restriction which permits only two travel directions. The extension of BML model by revising this restriction is worth further research.

In addition, we study the destination arrival rate and the convergence time. The arrival rate in C-2 falls faster than that in C-1 in the transition region except in the Scenario-3. We observe the convergence speed in C-1 is faster than that in C-2 in almost all cases except for the high density region in some instances. The convergence time rises faster in the high density region than the low density region.

In our simulation, we suppose all cars start out simultaneously on each way to their destinations. We know that this is not the case in actual traffic. It is possible the distribution of car departure time has a great effect on traffic. The departure time differences among all cars perhaps can't be ignored, which needs to be investigated further.

In our model, we assume that every site within the workplace can hold an infinite number of cars, which is impossible in actual traffic. The next step we will limit the capacity of site to accommodate cars and adopt more flexible route assignments, which will be closer to actual traffic.

\section*{Acknowledgments}
This work is financially supported by the National Key Technology R\&D Program (Grant No. 2009BAG13A05) and the Key Science and Technology Programs of Dongguan City (Grant No. 200910814006). We thank L. Li from the key 973 project at the ITS Laboratory of Tsinghua University for help and discussions. We are thankful to B. Leng at the School of Computer Science and Engineering of Beihang University (BUAA) for his instructing in academic writing. We also thank J. X. Zhang for setting up the computing environment for our simulations.

\end{document}